\documentclass[sigconf,nonacm]{acmart}
\usepackage{amsmath,amsfonts}
\usepackage{algorithmic}
\usepackage{graphicx}
\usepackage{textcomp}
\usepackage{xcolor}
\usepackage{url}
\usepackage{hyperref}
\usepackage{soul}
\usepackage{hyperref}
\usepackage{enumitem}
\usepackage{listings}
\usepackage{cleveref}
\usepackage{booktabs} 
\usepackage{xspace}
\usepackage{microtype}
\usepackage{subcaption}
\usepackage{balance}
\usepackage{mathtools}
\usepackage{scalerel}
\usepackage{tcolorbox}
\usepackage{microtype}
\usepackage[section]{placeins}
\usepackage[ruled]{algorithm2e}
\usepackage{algorithmic}
\usepackage{diagbox}
\usepackage{fancyvrb}
\usepackage{relsize}
\usepackage{xcolor}
\usepackage{pgfplots}
\usetikzlibrary{arrows.meta,
                arrows,
                chains,
                positioning,
                calc,
                shapes.geometric
                }
\usepackage{filecontents}
\usepackage[normalem]{ulem}
\usepackage{array}
\usepackage{multirow}
\usepackage{float}
\usepackage{caption}
\usepackage{booktabs}
\usepackage{threeparttable}
\usepackage{makecell}

\makeatletter
\newcommand{\scalefont}[1]{%
  \edef\scale@fontsize{\fpeval{#1*\f@size}}%
  \edef\scale@fontbaselineskip{\fpeval{1.2*\scale@fontsize}}%
  \fontsize{\scale@fontsize}{\scale@fontbaselineskip}\selectfont
}
\makeatother

\AtBeginDocument{%
  \providecommand\BibTeX{{%
    \normalfont B\kern-0.5em{\scshape i\kern-0.25em b}\kern-0.8em\TeX}}}

\begin{document}

\title{{\tt AMULET}: Adaptive Matrix-Multiplication-Like Tasks
}

\author{Junyoung Kim}
\affiliation{%
  \institution{Columbia University}
  \city{New York}
  \state{NY}
  \country{USA}
  \postcode{10027}}
\email{junyoung2@cs.columbia.edu}

\author{Kenneth Ross}
\affiliation{%
  \institution{Columbia University}
  \city{New York}
  \city{NY}
  \country{USA}
  \postcode{10027}}
\email{kar@cs.columbia.edu}

\author{Eric Sedlar}
\affiliation{%
  \institution{Oracle Labs}
  \city{Redwood City}
  \state{CA}
  \country{USA}
  }
\email{eric.sedlar@oracle.com}

\author{Lukas Stadler}
\affiliation{%
  \institution{Oracle Labs}
  \city{Linz}
  \country{Austria}
  }
\email{lukas.stadler@oracle.com}

\newcommand{\mmlt}{Amulet}
\DeclarePairedDelimiter\ceil{\lceil}{\rceil}

\makeatletter
\newcommand{\thickhline}{%
    \noalign {\ifnum 0=`}\fi \hrule height 1pt
    \futurelet \reserved@a \@xhline
}
\newcolumntype{"}{@{\hskip\tabcolsep\vrule width 1pt\hskip\tabcolsep}}
\makeatother

\newcolumntype{P}[1]{>{\centering\arraybackslash}p{#1}}

\begin{abstract}
Many useful tasks in data science and machine learning applications can be written as simple variations of matrix multiplication. However, users have difficulty performing such tasks as existing matrix/vector libraries support only a limited class of computations hand-tuned for each unique hardware platform. Users can alternatively write the task as a simple nested loop but current compilers are not sophisticated enough to generate fast code for the task written in this way. To address these issues, we extend an open-source compiler to recognize and optimize these matrix multiplication-like tasks. Our framework, called \mmlt{}, uses both database-style and compiler optimization techniques to generate fast code tailored to its execution environment. We show through experiments that \mmlt{} achieves speedups on a variety of matrix multiplication-like tasks compared to existing compilers. For large matrices \mmlt{} typically performs within 15\% of hand-tuned matrix multiplication libraries, while handling a much broader class of computations.
\end{abstract}

\keywords{Compilers, Data analytics}

\maketitle

\section{Introduction}\label{s:introduction}
Many useful data science tasks can be represented as simple variations of matrix multiplication. For example, consider a retail data analysis application where {\tt A[i][k]} records how many units of product k were bought by customer i in one year, and {\tt B[k][j]} records the list price of product k offered by seller j. The matrix product {\tt R[i][j]} shows how much customer i would have paid seller j if they had bought their products from that seller. Now suppose that sellers offer a discount (recorded as {\tt dis[j]}) if the total dollar sales of the item exceed a seller-dependent threshold (recorded as {\tt thres[j]}). Analysts could then write the code in Figure~\ref{fig:intro_task_example} to calculate {\tt R[i][j]}. Matrix multiplication-like tasks are also used in applications such as weather and climate prediction ~\cite{TensorcoreHPC} and earthquake simulation ~\cite{earthquake18ich}. In addition, many SQL queries using query operators such as natural joins~\cite{amossen09joinsparsematmul,deep20Fastjoinmatmul} and group-by aggregates can be translated to matrix multiplication~\cite{hu21tcudb}.

\begin{figure}
\begin{Verbatim}[frame=single, fontsize=\small]
for(i = 0; i < M; i++)
 for(j = 0; j < N; j++)
  for(k = 0; k < K; k++)
   R[i][j] += A[i][k]*B[k][j]-
   (A[i][k]*B[k][j]>thres[j])*A[i][k]*B[k][j]*dis[j];
\end{Verbatim}

\caption{Nested loop to calculate expected revenue}
\label{fig:intro_task_example}
\end{figure}

Similar code could be used in a machine learning (ML) application. Suppose {\tt A[i][k]} corresponds to the weight of observation i for feature k, and {\tt B[k][j]} is the strength of feature k at location j. The matrix product {\tt R[i][j]} is the net weighted strength for each observation i at location j. Suppose that analysts want to be more sophisticated, and want to separate out strong signals from many weak signals that add up and obscure the strong signal. In order to do this, the analyst decides to favor high single products by doubling the portion of the product that is higher than a threshold. The code for this task could be written as shown in Figure~\ref{fig:intro_ml}.


\begin{figure}
\begin{Verbatim}[frame=single, fontsize=\small]
for(i = 0; i < M; i++)
 for(j = 0; j < N; j++)
  for(k = 0; k < K; k++)
   R[i][j] += A[i][k]*B[k][j]+
   (A[i][k]*B[k][j]>thres[j])*(A[i][k]*B[k][j]-thres[j]);
\end{Verbatim}
\caption{Nested loop to calculate net weighted strength}
\label{fig:intro_ml}
\end{figure}

These data science tasks are easily represented by writing code as a nested for loop. The problem is that doing so likely results in bad performance as current compilers are not sophisticated enough to produce fast machine code tailored for every type of execution environment. Due to varying hardware characteristics such as cache sizes, number of cores, and available instructions, performance diversity may happen as code that is optimal for some system may not be optimal for another system. Even on the same system, the optimal code may differ for each execution when applications share resources, such as in the context of cloud computing, datacenters ~\cite{zhao18tilingdatacenters} or UDFs in queries. Furthermore, complex code in a loop challenges compilers, which need to perform some code analysis to verify that transformations preserve the semantics of the program. The burden on the compiler continues to increase as more data science tasks can be done entirely in main memory as the amount of RAM available to modern hardware increases ~\cite{Faerber2017}. With disk I/O no longer being the bottleneck, the speed of the generated machine code becomes crucial to performance.

Alternatively, users can turn to existing matrix/vector libraries that are manually and individually tuned for a wide range of hardware platforms to execute matrix operations. However, these libraries are not flexible enough to represent matrix multiplication-like computations such as the previously shown examples. They also do not account for dynamically changing execution environments on the same hardware.


Due to these issues, these matrix multiplication-like tasks are not employed in the machine learning research community as they are slow and do not scale\textcolor{black}{~\cite{VondrickCommunication}}. If they were quick to execute, they would be considered more in research, and could lead to the discovery of new ML models and more useful matrix mutliplication-like tasks~\cite{VondrickCommunication, barham19ml}. Another positive impact is that it would help spread adoption of ML as it would enable ML students write simple code that is also fast. ML students currently often write correct code that is slow to execute because they encode the linear algebra operations directly rather than calling libraries. Introducing a compiler that allows fast execution of simply written ML tasks will make code more readable and reduce bugs. 

A tuned compiler for matrix-multiplication-like tasks could also be used in the context of a DBMS, where an SQL query optimizer could rewrite the inner loop(s) of an SQL query using a nested loops formulation. Invoking the compiler on the rewritten query would achieve good performance for SQL queries or subqueries encoding matrix multiplication-like tasks.

In this paper, we propose \mmlt{},\footnote{Adaptive MUltiplication-LikE Tasks} a compiler framework that automatically recognizes and optimizes matrix multiplication-like data science tasks. We recognize the opportunity to use database query optimization techniques in the field of compilers ~\cite{thriving2019pirk} and build upon Vectorwise DBMS's~\cite{Raducanu:2013:MAV} dynamic query execution scheme and the compiler proposed by Zhang et al. ~\cite{wangdaacg2020} to generate fast code for a variety of execution environments. We implement our techniques as an extension to an open source compiler (Graal ~\cite{Wurthinger:2013} and Truffle~\cite{wimmer2012truffle}). After recognizing that a loop written by the programmer is a matrix multiplication-like task, our compiler transforms and compiles the loop into a parameterizable tiled loop tailored for matrix multiplication. The parameterized code is then experimented with different optimization parameters at runtime using a relatively small portion of the input data, and based on runtime feedback the fastest parameter configuration is used to execute on the remaining input data. Additionally, we also define a simple, declarative programming syntax that a programmer can use to easily specify a task over arrays. 

We show through experiments that \mmlt{} achieves speedups on a wide range of matrix multiplication-like tasks compared to code generated by existing compilers, and as a side benefit gets matrix multiplication performance for large matrices within 15\% of manually tuned libraries. We also show that the overhead due to compilation is manageable compared to existing compilers, and is negligible compared to the execution time of the program when operating on large matrices.

The remainder of the paper is structured as follows. In Section~\ref{fast matmul}, we provide a brief summary of how fast matrix multiplication programs are written for modern hardware. In Section~\ref{s:adaptive code generation}, we explain the techniques used by \mmlt{} for adaptively generating optimized code. Section~\ref{s:implementation} describes how we implemented our techniques in Graal and Truffle. Experimental evaluation on a variety of queries and matrix sizes is done in Section~\ref{s:experiment results}.

\section{Background and Related Work}


To get around the problems caused by performance diversity on different hardware, the Vectorwise DBMS uses a query processing scheme where the performance of multiple, precompiled, different implementations of the same operator are tested out at query runtime, so that the fastest operator can be used most of the time ~\cite{Raducanu:2013:MAV}. This approach allows for (a) less reliance on complex and potentially inaccuarate cost modeling, (b) fast performance as most of the data will be executed using the fastest query operator, and (c) adaptivity to changing execution environments.

However, the Vectorwise system is limited in several ways. First, its approach has limited usage outside of in-memory relational DBMSs. A second limitation is that the system only supports adaptivity between precompiled code fragments for the essential DBMS operators, so user-defined code is not considered. Finally, the underlying compiler is treated as a black-box, which leads to unpredictable performance depending on what compiler is used with which compiler settings.


Building upon and addressing the limitations of the approach taken by Vectorwise, Zhang et al.\ extended an open-source compiler with functionality that automatically detects performance diversity possibilities in a for loop to generate several execution plans ~\cite{wangdaacg2020}. These plans are run on several chunks of data at a time, and runtime feedback is used to determine the fastest plan, which is employed for an extended number of data chunks. This process of testing and executing with the best plan is repeated several times during the execution of a loop, so that the system can adaptively switch between plans at runtime.
The advantages of this compiler can be summarized as follows: (a) dynamic, user-defined queries are supported as code is compiled at query-time; (b) database-style and compiler optimization co-exist, eliminating some of the mismatches that happen when the compiler is used as a black-box by a DBMS; (c) the system achieves good performance by dynamically responding to changes in the execution environment, by choosing the best plan for the situation.

However, the approach of Zhang et al.~\cite{wangdaacg2020} is limited in that it only supports queries that can be written using a single for loop, which limits its usage for other tasks that involve nested for loops. Another limitation is that the compiler does not optimize memory access patterns, which could lead to poor cache locality. We aim to address these issues by extending a compiler to support adaptive memory access optimizations for matrix multiplication-like tasks that can be written in a simple, declarative programming syntax.


There are various works on how to optimize dense matrix multiplication on CPUs ~\cite{GotoMat08, vanzee15blis, su17autogenblas}. However, their performance relies on specific parameters found before runtime and thus do not apply well on changing execution environments. \textcolor{black}{~\cite{dinh20communication} uses a mathematical model to find optimal tile sizes for nested loop tasks including matrix multiplication given machine characteristics such as cache size. However, its results are theoretical, and it is unknown if its methods account for all factors that affect performance, which limits its use in modern ML compilers such as TVM~\cite{chen2018tvm}.} ~\cite{faingnaert22gpugemm} allows easy implementation of variants of matrix multiplication on GPUs, but finding optimal parameters for execution is still left to the programmer. Spiral ~\cite{franchetti2018spiral} automates the creation of high-performance libraries for commonly used computations including matrix multiplication. However, a specific API must be used to express computations, and hardware-specific optimizations are done once before runtime.

Adaptive Query Processing ~\cite{deshpande2007adaptive, avnur2000eddies} is a technique where the database aims to refine a query plan at runtime based on the statistics gathered during the intermediate stages of query execution. ~\cite{PalkWeld18} proposes a programming language that uses adaptive query processing techniques during execution. It differs from our work in that it does not target matrix multiplication-like tasks, and requires the user to represent computations through a class defined by the library. ~\cite{KohnL018} introduces a method that adaptively chooses how much to compile the query, and ~\cite{MenonPM20} reduces the burden of compiling several query plans by making compiled queries permutable so that they can be changed during execution.

Compiler loop vectorization techniques ~\cite{allen1987auto, nuzman2008outer} aim to exploit data-level parallelism to speedup computation kernels. SIMD optimizations have also been applied in a database context to speed up various operators such as joins, sorting, scans and compression ~\cite{B13, C08, ZR02, L16, P15}.

There are several execution environment-aware optimization techniques in the compiler field. ~\cite{clint98atlas, FrigoFFTW98} use autotuning, where different implementations for code are tested offline and the fastest implementation is saved and is used on subsequent runs. This differs from \mmlt{}, where adaptive optimizations happen during runtime. ~\cite{DinizDynamic1997} introduces a technique where the best synchronization scheme for a parallel program is found by trying out several different synchronization schemes at runtime. ~\cite{TavaTile11, srinivas15reactive, zhao18tilingdatacenters} aims to find the optimal tile sizes for a tiled nested loop at runtime.

Compilers specifically for ML have been recently developed for the similar goal of automatically generating fast, optimized code for a diverse range of hardware while presenting an easy way for the user to write code for tensor/matrix operations. ~\cite{XLA} generates model-specific computation kernels for Tensorflow~\cite{abadi2016tensorflow} by fusing operations into a single kernel. ~\cite{vasilache2018tensor} proposes a concise domain-specific language for expressing tensor operations. Both compilers support autotuning. TVM ~\cite{chen2018tvm} aims to generate code for various hardware by training a ML-based cost model to evaluate code during an offline autotuning phase where the user defines the exploration space for optimizations. For operations like matrix multiplication, TVM also allows the user to manually configure optimizations such as blocking and vectorization, but register level optimizations such as those described by Goto et al ~\cite{GotoMat08} are not considered. ~\cite{taso2019jia} automatically generates graph substitutions for deep neural networks represented as a graph. ~\cite{yang2016systematic} aims to find energy-efficient blockings for CNN-like computations written in a nested loop by using an analytical model to estimate energy needed for the computation.

Multiple works on exploring linear algebra operations in the context of DBMSs have been proposed. ~\cite{YuanTen21} introduces a variant of relational algebra to be used as the backend of a ML system that operates on tensors. ~\cite{LuoAuto21} proposes a framework for optimizing matrix operations on DBMSs by representing matrix operations as a graph and then using a cost model to explore and find the optimal implementation.~\cite{jankov2019decrec} adds for-loop like syntax to SQL to support ML tasks on a RDBMS. ~\cite{geerts2021express} extends a matrix query language so that it can express classic linear algebra algorithms. ~\cite{arrayql2022} integrates a declarative language for processing arrays into an RDBMS.


\section{Fast matrix multiplication} \label{fast matmul}

We briefly describe the algorithm by Goto et al. ~\cite{GotoMat08}, the current state-of-the-art algorithm for matrix multiplication on CPUs. Goto's algorithm is widely used in common linear algebra libraries such as Intel MKL ~\cite{mkl} and OpenBLAS ~\cite{OpenBLAS}.

\begin{figure}
    \centering
    \includegraphics[scale=0.26]{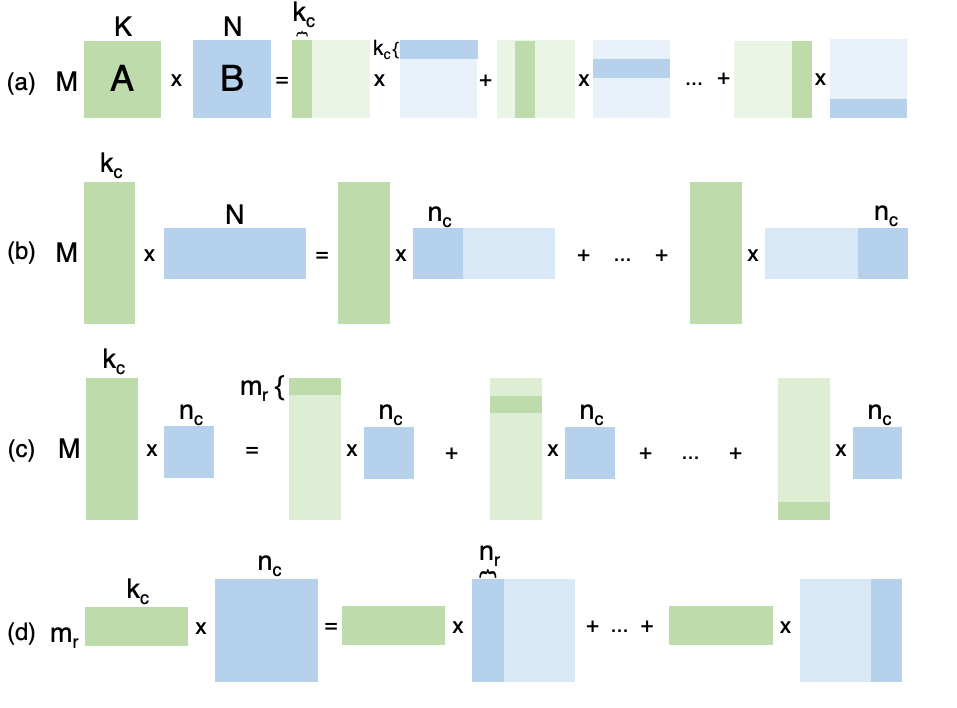}
    \caption{Processing of partitioning matrix multiplication into smaller subtasks}
    \label{fig:matmul partitioning}
\end{figure}

Given that we are multiplying two matrices A and B where A has dimensions $M \times K$ and B has dimensions $K \times N$, and the results are stored in matrix C of dimension $M \times N$, the process of matrix multiplication is divided into multiple, smaller subtasks. A matrix multiplication is first divided into the sum of $K / k_c$ individual panel-panel matrix multiplications (Figure~\ref{fig:matmul partitioning}a). Each panel-panel multiplication is then split into $N / n_c$ panel-block multiplications. (Figure~\ref{fig:matmul partitioning}b). Furthermore, each panel in a panel-block multiplication can be divided into slices of size $m_r \times k_c$ (Figure~\ref{fig:matmul partitioning}c). \textcolor{black}{We want} to make each block of size $k_c \times n_c$ fit in the L2 cache while being as big as possible, so that the cost of reading each block into memory is amortized over the computation with the left panel, where we stream over each slice of size \textcolor{black}{$m_r \times k_c$} and bring it into memory to multiply with the block.

Furthermore, each block in a slice-block multiplication can be divided into slices of size $k_c \times n_r$ (Figure~\ref{fig:matmul partitioning}d). \textcolor{black}{Similar to the previous process,} we want each slice of size $m_r \times k_c$ from A to fit in the L1 cache while being as big as possible, so that the cost of reading each slice into memory is amortized over the computation with the right block, where we stream over each slice of size \textcolor{black}{$k_c \times n_r$} from B.

We can optimize slice-slice multiplication by using SIMD instructions provided on modern hardware to operate on vectors at a time, thus reducing the total number of instructions needed to do a slice-slice multiplication.

\begin{figure}
\begin{Verbatim}[commandchars=\\\{\}, frame=single]
inner_kernel_12x16(double A[M][K], double B[K][N], 
    double C[M][N], int k_c, int i, int k, int j) \{

    simd_reg a_reg;
    simd_reg b_reg0;
    simd_reg b_reg1;
    simd_reg c00, c01, c10, c11, c20, c21, c30, c31,
            ... c90, c91, c100, c101, c110, c111;
    
    \textcolor{green!45!black}{// zero all registers to store subresults}        
    simd_setzero(c00);
    simd_setzero(c01);
    ...
    simd_setzero(c111);
    
    for(int kk = 0; kk < k_c; k++) \{
        \textcolor{green!45!black}{// load B[k+kk][j]~B[k+kk][j+7] to b_reg0}
        b_reg0 = simd_load(&B[k+kk][j]);
        b_reg1 = simd_load(&B[k+kk][j+8]);
        
        \textcolor{green!45!black}{// calculate subresults}
        
        \textcolor{green!45!black}{// broadcast value of A[i][k+kk] to a_reg}
        a_reg = simd_broadcast(&A[i][k+kk]);
        c00 = simd_fmadd(a_reg, b_reg0, c00);
        c01 = simd_fmadd(a_reg, b_reg1, c01);
        ...
        a_reg = simd_broadcast(&A[i+11][k+kk]);
        c110 = simd_fmadd(a_reg, b_reg0, c110);
        c111 = simd_fmadd(a_reg, b_reg1, c111);
    \}
    
    \textcolor{green!45!black}{// add subresults to C}
    c00 = simd_add(&C[i][j], c00);
    simd_store(&C[i][j], c00);
    c01 = simd_add(&C[i][j+8], c01);
    simd_store(&C[i][j+8], c01);
    ...
    c111 = simd_add(&C[i+11][j+8], c111);
    simd_store(&C[i+11][j+8], c111);
\}
\end{Verbatim}
\caption{12x16 Inner kernel function}
\label{fig:12x16 inner kernel}
\end{figure}

Finally, we can optimize slice-slice multiplication even further by doing register blocking. Since a single element of A is multiplied with multiple elements of B and vice versa, we can reduce the number of loads to SIMD registers from memory by keeping loaded values in registers as long as possible, i.e., until the values in the register are multiplied with all values that it needs to multiplied with. \textcolor{black}{Additionally, we can reduce the number of stores from SIMD registers to memory by accumulating matrix multiplication subresults in idle SIMD registers and writing them out to the result matrix at the end of slice-slice multiplication, as opposed to writing matrix multiplication subresults directly to the result array after every multiplication.} These two optimizations reduce the number of stores and loads from/to registers, thus reducing the overhead due to moving data between the CPU and cache. Where $S_w$ is the width of a SIMD register (in terms of matrix elements it can hold), for register blocking to be possible, we need $m_r \times \ceil{(n_r / S_w)} + 1 + \ceil{(n_r / S_w)}$ SIMD registers, as $m_r \times \ceil{(n_r / S_w)}$ registers are needed to hold $m_r \times n_r$ subresults, 1 register is needed to continuously load elements from the slice from A, and $\ceil{(n_r / S_w)}$ registers are needed to load rows from the slice from B to multiply with elements from A. So where $R$ is the number of available SIMD registers, to perform register blocking the following inequality must hold.

\[m_r \times \ceil{(n_r / S_w)} + 1 + \ceil{(n_r / S_w)} \leq R\]

The high performance function for computing $m_r \times n_r$ subresults given two slices of A and B is referred to as the \textbf{inner kernel}. An example inner kernel written in pseudocode for computing $12 \times 16$ subresults on matrices containing double precision numbers is shown in Figure~\ref{fig:12x16 inner kernel}. In general the kernel size is chosen so that it uses the maximum number of registers, as doing so increases the ratio of \emph{number of calculated subresults : reads from memory}, thus resulting in a more efficient inner kernel.

In summary, the code for achieving fast matrix multiplication is shown in Figure~\ref{fig:optimized matmul code}, where the function inner\_kernel() does a slice-slice matrix multiplication using SIMD and register blocking, where the top left corner of the slice in A is located at $(i, k)$, the top left corner of the slice in B is located at $(k,j+jj)$, and the length/height of the slices in A and B respectively is $k_c$.

\begin{figure}[h!]
\begin{Verbatim}[frame=single]
for(k = 0; k < K; k+=k_c)
 for(j = 0; j < N; j+=n_c)
  for(i = 0; i < M; i+=m_r)
   for(jj = 0; jj < n_c; jj+=n_r)
    inner_kernel(A, B, C, k_c, i, k, j+jj);
\end{Verbatim}
\caption{Optimized matrix multiplication code}
\label{fig:optimized matmul code}
\end{figure}

Since cache sizes vary for each machine, the optimal values of $k_c$ and $n_c$ are dependent on the execution environment, which can lead to performance diversity as the optimal matrix multiplication code for one machine may be slow on another machine due to poor cache utilization.

\subsection{Additional techniques used by libraries} \label{libraryadditionaltechniques}

Goto et al. ~\cite{GotoMat08} also explains how performance can further be improved by copying the matrices into a format where the matrix elements are ordered in memory in the same order they would be accessed during matrix multiplication. This technique, called packing, ensures that all submatrices that occur due to the partitioning process shown in Figure~\ref{fig:matmul partitioning} are contiguous in memory, which minimizes the number of TLB entries needed when performing a matrix multiplication subtask, which in turn minimizes the number of TLB misses that happen during matrix multiplication. Common libraries such as Intel MKL ~\cite{mkl} and OpenBLAS ~\cite{OpenBLAS} use packing. Libraries can also further optimize the inner kernel by using {\tt movddup} and {\tt broadcastf32x4} SIMD instructions to reduce the number of instructions needed in the inner kernel, at the cost of using more SIMD registers. This optimization is only possible if packing is used. A compiler should not assume that a user would want it to allocate extra memory to store the input data redundantly, even if that storage has better access properties. Nevertheless, we can in principle support a compiler optimization option that would allow a user to specify that such auto-packing is allowed. Our experimental results show that with packing enabled, we obtain performance very close to that of high-performance libraries.


\section{Adaptive Code Generation}\label{s:adaptive code generation}

In this section we describe how we generate adaptive code for matrix multiplication-like tasks.

\begin{figure}[H]
\centering
\begin{tikzpicture}[>=latex']
        \tikzset{rblock/.style= {draw, rectangle, align=center,minimum width=1.5cm,minimum height=1.2cm},
        input/.style={ 
        draw,
        trapezium,
        trapezium left angle=60,
        trapezium right angle=120,
        minimum width=2.2cm,
        align=center,
        minimum height=1cm
    },
        }
        \node [rblock]  (start) {Task \\ recognition};
        \node [rblock, right =0.5cm of start] (acquire) {Parameterized \\ code generation};
        \node [rblock, right =0.5cm of acquire] (rgb2gray) {Adaptive \\ execution};

        \path[draw,->] (start) edge (acquire)
                    (acquire) edge (rgb2gray)
                    ;
\end{tikzpicture}
\caption{Process Overview}
\label{fig:proces overview}
\end{figure}
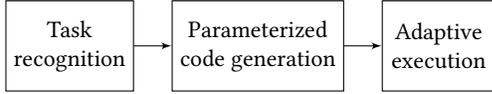

An overview of \mmlt{}'s process to generate and execute fast code for matrix multiplication-like tasks is shown in Figure~\ref{fig:proces overview}. At \textbf{task recognition}, \mmlt{} recognizes that a loop is a matrix multiplication-like task. During \textbf{parameterized code generation,} code for the task that can be run with a variety of optimization parameters is generated. Finally, in the \textbf{adaptive execution stage,} various configurations of optimization parameters are tested on a small subset of the input data to find a configuration that yields fast performance, and the remainder of the input data is executed using the fastest configuration. We will go over each stage in detail in the following sections.

\vspace{-0.0cm}
\subsection{Declarative Syntax}\label{s:declarative syntax}

We first explain the syntax of loops \mmlt{} will target. We define a new declarative syntax for expressing data science tasks that are often written as tight nested loops in an imperative programming language.

The syntax is shown in Figure~\ref{fig:acg_decl_syntax}. The code means that the innermost loop statement is to be executed over every combination of values of loop variables $var_1, var_2, ..., var_n$, where $var_1$ is in range $[s_1, e_1)$, $var_2$ is in range $[s_2, e_2)$, and so on. This syntax assumes that the order of the iterations in which the loop is executed does not affect the final result. As an example, the declarative form of the loop in Figure~\ref{fig:intro_task_example} is shown in Figure~\ref{fig:acg_revenue_decl}.

\vspace{-1mm}
\begin{figure}[H]
\begin{Verbatim}[frame=single, fontsize=\small]
where(var1 in [s1..e1] and ... and varN in [sN..eN]) {
    // computation statement
} 
\end{Verbatim}
\vspace{-1.5mm}
\caption{Declarative form of loops}
\label{fig:acg_decl_syntax}
\end{figure}

\vspace{-5mm}
\begin{figure}[H]
\begin{Verbatim}[frame=single, fontsize=\small]
where(i in [0..M] and j in [0..N] and k in [0..K]) {
    R[i][j] += A[i][k]*B[k][j]-
     (A[i][k]*B[k][j]>thres[j])*A[i][k]*B[k][j]*dis[j];
} 
\end{Verbatim}
\vspace{-1.5mm}
\caption{Declarative form of revenue query}
\label{fig:acg_revenue_decl}
\end{figure}


The advantages of this syntax are twofold.

\begin{itemize}
    \item It allows an easier way for programmers to write data-intensive queries in the context of a conventional programming language.
    \item It signals to the compiler where to apply adaptive optimization techniques, and certifies that loop reordering is safe without complex code analysis.
\end{itemize}

\textcolor{black}{To preserve compatibility with old code, users still have the option to write loops in the original, nested format, and can signal to \mmlt{} to perform adaptive execution by writing a \textit{pragma} ``amulet'' directly above the for-loop they wish to optimize.}

\subsection{Matrix Multiplication-like Task Recognition}\label{s:matmul task conditions}

After \mmlt{} identifies a loop is written in declarative form, it checks whether the loop is a matrix multiplication-like task before generating adaptive code. A loop is a matrix multiplication-like task if the following conditions are met.\footnote{If the conditions are not met, the code is rewritten using nested loops and processed conventionally.}

\begin{enumerate}
    \item The computation can be written as a single statement.
    \item The computation statement is iterated using three loop variables, let's say i, j, and k.
    \item The computation statement is an assignment or an increment to a 2D array indexed using two of the loop variables. Without loss of generality, let's say it is indexed using i and j (e.g. $R[i][j]$).
    \item The computation statement's right hand side indexes two distinct 2D arrays, and the arrays are indexed using i, k, and k, j respectively. (e.g. $A[i][k], B[k][j])$)
\end{enumerate}

While the computation statement could have also been written using {\tt if} statements or the ternary operator (e.g. \verb# a<b ? E1 : E2#), both of which our compiler currently does not recognize, we could easily handle these cases with a compiler pre-processing step to rewrite the code into a form that satisfies the conditions above.

\subsection{Parameterized Code Generation}

After the compiler recognizes that a loop is a matrix multiplication-like task, the compiler has to transform the declarative loop into a paramaterized loop suited for matrix multiplication as shown in Section \ref{fast matmul}. To do this, code for the inner kernel is first generated.

Unlike matrix multiplication, where code for the inner kernel is relatively straightforward, in the case of general matrix multiplication-like tasks \mmlt{} has to 1) generate efficient assembly code for calculating subresults, and 2) determine the appropriate kernel size.

There is a tradeoff between a large inner kernel size and efficient assembly code. Having a large inner kernel increases the efficiency of register blocking as more subresults are calculated for each element loaded into a register. On the other hand, since more registers are used to store intermediate subresults, more instructions, and thus less efficient assembly code may be generated as the compiler must generate code with a limited register budget. Using a small inner kernel size reduces the efficiency of register blocking, but may result in more efficient assembly code to calculate subresults as the compiler has more free registers to work with.

Generating code for the inner kernel is divided into three parts. We first generate pseudo assembly instructions for computing a single subresult to check how many extra registers our code needs. Second, the compiler determines the appropriate kernel size based on the number of registers calculated from the previous step. Finally, the compiler generates actual assembly code for the inner kernel based on the pseudo instructions and kernel size calculated from the previous steps.

\subsubsection{Generating pseudo instructions for calculating subresults}\label{s:generating pseudo insts}

To calculate a single subresult, the computation statement in the declarative loop is first parsed into a tree, and then common subexpression elimination is performed to transform the tree into a directed acyclic graph. 

\begin{figure}
    \centering
    \includegraphics[scale=0.3]{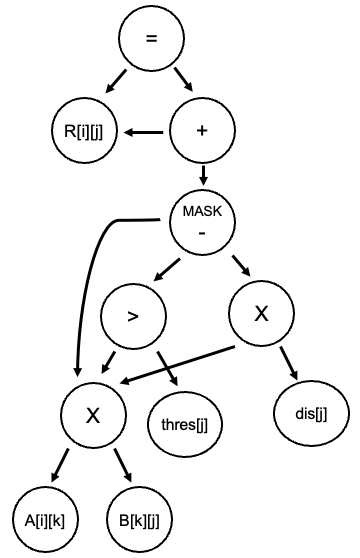}
    \caption{Expression DAG of query shown in Figure~\ref{fig:acg_revenue_decl}}
    \label{fig:expr_dag}
\end{figure}

The compiler's goal is to then traverse the expression DAG to produce instructions that use the fewest registers. It is known that the problem generating an evaluation of an expression DAG that is optimal (i.e. uses the fewest registers) is NP-complete~\textcolor{black}{~\cite{sethi1973complete}}. So as an alternative we use the heuristic \textit{labelfs}~\cite{kebler91arandomized} to traverse the DAG and generate assembly code, as it has been shown to produce fairly good results compared to the optimal code. At this stage in compilation we assume that the values of the leaves are already stored in (SIMD) registers, as the point of the inner kernel is to reduce loads to SIMD registers by keeping loaded values in SIMD registers as long as possible. We also assume that we have an infinite number of SIMD registers, and allocate a new SIMD register whenever it is absolutely necessary to do so (i.e. all previously allocated SIMD registers are still holding a live variable). 

As an example, the expression DAG for the query in Figure~\ref{fig:acg_revenue_decl} is shown in Figure~\ref{fig:expr_dag}. `Mask -' is an operation where the result is \textbf{Child1 - Child2 $\times$ Child3}. This corresponds to a SIMD operation where the results of evaluating a boolean expression are used to mask a SIMD subtraction. An example translation of this expression DAG into pseudo instructions is shown in Figure~\ref{fig:pseudo_instructions}. A, B, R, THRES, DIS are placeholder variables that correspond to the values in matrices A, B, R and arrays {\tt thres} and {\tt dis} that are used to compute a subresult, and REG, MASKREG indicate SIMD registers and mask registers used to store intermediate values while computing a subresult.


\begin{figure}
\begin{Verbatim}[frame=single]
mul REG0 A B
gtcmp MASKREG0 REG0 THRES
mul REG1 REG0 DIS
masksub REG0 MASKREG0 REG0 REG1
add R R REG0
\end{Verbatim}
\caption{Pseudo instructions for DAG in Figure~\ref{fig:expr_dag}}
\label{fig:pseudo_instructions}
\end{figure}

We can determine how many extra SIMD registers we need to perform the computation by checking how many additional registers the compiler allocated. \textcolor{black}{For the expression DAG shown in Figure~\ref{fig:expr_dag}, two extra registers would be needed ({\tt REG0} and {\tt REG1} shown in Figure~\ref{fig:pseudo_instructions}).}

\subsubsection{Choosing Kernel Size}\label{s:choosing kernel size}

The SIMD registers that are needed for an inner kernel of size $I_h \times I_w$ are the following. On a matrix multiplication-like task where the first matrix is indexed using loop variables $(i,k)$ and the second matrix is indexed using loop variables $(k, j)$, and where $S_w$ is the width of a SIMD register in terms of matrix elements it can hold,

\begin{enumerate}
    \item $I_h \times \ceil{(I_w / S_w)}$ registers to hold subresults
    \item 1 register needed to load elements from A
    \item $\ceil{(I_w / S_w)}$ registers are needed to load rows from B
    \item For each leaf node in the expression DAG excluding nodes corresponding to $A$, $B$ and the result matrix,
    \begin{enumerate}
        \item If the value is a constant, 1 register to hold the constant throughout the computation
        \item If the value is an array indexed with $i$, 1 register needed to load elements from it (in a similar fashion to A)
        \item If the value is an array indexed with $j$, $\ceil{(I_w / S_w)}$ registers needed to load rows from it (in a similar fashion to B)
        \item If the value is an array indexed with both $i, j$, 1 register needed to continuously load elements from it (needed to be loaded for every subresult calculation)
    \end{enumerate}
    \item Additional registers allocated by the compiler when generating pseudo instructions for calculating a subresult
\end{enumerate}

\mmlt{} then chooses the appropriate kernel size for the inner kernel function by choosing the largest size that does not exceed the register budget. To calculate this, the compiler starts from the largest inner kernel size possible where $I_h = 12$ and $I_w = 16$ and decreases $I_h$ until the total number of SIMD registers needed is within the system's register budget. The largest inner kernel size possible is $12 \times 16$ as opposed to $14 \times 16$ as we use an optimization where the addresses of the rows of A (i.e. {\tt \&A[i], \&A[i+1], ...}) are stored in general purpose registers before the computation loop in order to reduce the number of instructions needed to access elements of A. We found that we had 12 spare general purpose registers to use for this purpose after taking into account general purpose registers that were already in use, for example the loop counter. So the largest inner kernel size is $12 \times 16$.

For the example query in Figure~\ref{fig:acg_revenue_decl} that operates on matrices of double-precision floating point numbers on a machine that has 32 512-bit SIMD registers, the compiler would calculate that a $12 \times 16$ size inner kernel would need

\begin{enumerate}
    \item 24 registers to hold subresults
    \item 1 register needed to load elements from $A$
    \item 2 registers needed to load rows from $B$
    \item 2 registers needed to load rows from {\tt thres[j]}, and 2 registers needed to load rows from {\tt dis[j]}
    \item 2 additional registers needed for generating instructions for calculating subresults ({\tt REG0} and {\tt REG1} shown in Figure~\ref{fig:pseudo_instructions}).
\end{enumerate}

\noindent
which is a total of 33 SIMD registers. Since this is larger than the number of registers usable by the machine, the calculation is repeated for an inner kernel size of $11 \times 16$, which requires 31 SIMD registers. \mmlt{} would then choose a kernel size of $11 \times 16$.

\subsubsection{Inner Kernel Code Generation}\label{s:inner kernel generation}

Using the pseudo instructions generated from the first step and the kernel size calculated from the second step, \mmlt{} now generates assembly code for the inner kernel. The format is the same as the inner kernel shown in Figure~\ref{fig:12x16 inner kernel}, except data corresponding to the leaves in the expression DAG are loaded into registers in addition to the data in A and B, and the instructions corresponding to the pseudo instructions for calculating subresults are generated.

In order to generate SIMD instructions from pseudo instructions, \mmlt{} generates SIMD equivalent instructions for each pseudo instruction (e.g. generate a SIMD add for {\tt{add}}), and placeholder variables are replaced with the SIMD register that contains the appropriate value needed to calculate the subresult.

The pseudo code for the inner kernel for the query shown in Figure~\ref{fig:acg_revenue_decl} is shown in Figure~\ref{fig:revenue inner kernel}. In the actual implementation the inner kernel is generated directly into assembly instructions and not in the form of a function written in a high level programming language.

\begin{figure}
\begin{Verbatim}[commandchars=\\\{\}, frame=single]
inner_kernel(double A[M][K], double B[K][N], 
    double C[M][N], int k_c, int i, int k, int j,
    double thres[N], double dis[N]) \{
    ...
    \textcolor{green!45!black}{//declare registers,}
    \textcolor{green!45!black}{//zero all registers to store subresults}        
    
    for(int kk = 0; kk < k_c; k++) \{
      b_reg0 = simd_load(&B[k+kk][j]);
      b_reg1 = simd_load(&B[k+kk][j+8]);
      thres_reg0 = simd_load(&thres[j]);
      thres_reg0 = simd_load(&thres[j+8]);
      dis_reg0 = simd_load(&dis[j]);
      dis_reg1 = simd_load(&dis[j+8]);
        
      \textcolor{green!45!black}{// calculate subresults}
        
      a_reg = simd_broadcast(&A[i][k+kk]);
      reg0 = simd_mul(a_reg, b_reg0);
      mask_reg0 = simd_gt_cmp(reg0, thres_reg0);
      reg1 = simd_mul(reg0, dis_reg0);
      reg0 = simd_mask_sub(reg0, mask_reg0, reg1);
      c00 = simd_add(c00, reg0);
      ...
    \}
    
    \textcolor{green!45!black}{// add subresults to C}
    ...
\}
\end{Verbatim}
\caption{Inner kernel function for query in Figure~\ref{fig:acg_revenue_decl}}
\label{fig:revenue inner kernel}
\end{figure}

\subsection{Adaptively Finding Optimal Parameters}\label{adaptive_param_strategy}

Once the inner kernel function is generated, the compiler creates a parameterized nested loop shown in Figure~\ref{fig:parameterized matmul code} to repeatedly invoke the inner kernel function over the two input matrices to produce the final result. In the code $k_c, n_c,$ loop\_start0, loop\_end0, loop\_start1, loop\_end1, loop\_start2, loop\_end2 are all variables that can be set before calling the nested loop, and $i_h, i_w$ are constants that correspond to the height and width of the inner kernel. In this section we show how the compiler adaptively finds the optimal values for $k_c$ and $n_c$ during runtime.

\begin{figure}
\begin{Verbatim}[frame=single]
for(k = loop_start0; k < loop_end0; k+=k_c)
 for(j = loop_start1; j < loop_end1; j+=n_c)
  for(i = loop_start2; i < loop_end2; i+=i_h)
   for(jj = 0; jj < n_c; jj+=i_w)
    inner_kernel(A, B, C, k_c, i, k, j+jj);
\end{Verbatim}
\caption{Parameterized matrix multiplication code}
\label{fig:parameterized matmul code}
\end{figure}

To execute the matrix multiplication-like task, $k_c$, $n_c$ and the loop\_start, loop\_end variables are set to some values before calling the nested loop. This allows \mmlt{} to calculate portions of the matrix multiplication-like task independently with different values of $k_c$ and $n_c$, thus allowing \mmlt{} to test different parameters before settling on values of $k_c$ and $n_c$ that yield the fastest execution time. These intermediate computations also all contribute to the final result. We devised a strategy that first finds the optimal value of $k_c$, and then finds the optimal value of $n_c$. This is because given a kernel size of $m_r \times n_r$, we want to find the biggest value of $k_c$ where $m_r \times k_c$ matrix elements fit in the L1 cache, and the biggest value of $n_c$ where $k_c \times n_c$ matrix elements fit in the L2 cache. So it makes sense to find the optimal value of $k_c$ first before searching for the optimal value of $n_c$.

\begin{table}[tb!]
\caption{AB 4096x4096 Matrix Multiplication Performance for Different Parameters \textcolor{black}{on Two Machines} (Unit: Seconds)}
\label{tab:ab_matmul_param}
\footnotesize
\def\arraystretch{1.1}%
\setlength{\tabcolsep}{4pt}
\begin{tabular}{|l|l|l|l|l|l|l|l|l|l|}
\hline
\backslashbox{KC}{NC} & 16 & 32 & 64 & 128 & 256 & 512 & 1024 & 2048 & 4096 \\ \hline
16 & 8.93 & 8.39 & 6.59 & 5.94 & 5.59 & 5.36 & 5.18 & 5.08 & 5.01\\ \hline
32 & 5.85 & 5.64 & 4.77 & 4.35 & 4.13 & 4.03 & 4.02 & 4.09 & 4.30\\ \hline
64 & 4.34 & 4.08 & 3.80 & 3.95 & 4.03 & 4.06 & 4.09 & 4.12 & 4.27\\ \hline
128 & 3.46 & 3.32 & 3.23 & 3.69 & 3.75 & 3.75 & 3.75 & 3.74 & 3.74\\ \hline
256 & 3.10 & 2.96 & 2.93 & 3.52 & 3.54 & 3.53 & 3.53 & 3.53 & 3.52\\ \hline
512 & 2.84 & 2.78 & 2.83 & 3.44 & 3.41 & 3.40 & 3.40 & 3.39 & 3.38\\ \hline
1024 & 2.90 & 2.80 & 2.88 & 3.48 & 3.40 & 3.35 & 3.33 & 3.34 & 4.13\\ \hline
2048 & 3.23 & 3.17 & 3.47 & 3.54 & 3.41 & 3.35 & 3.33 & 4.49 & 7.53\\ \hline
4096 & 4.39 & 4.22 & 4.00 & 3.74 & 3.59 & 3.52 & 4.74 & 8.13 & 8.48\\ \hline
 \multicolumn{9}{c}{} \\
\end{tabular}
\textcolor{black}{
\begin{tabular}{|l|l|l|l|l|l|l|l|l|l|}
\hline
\backslashbox{KC}{NC} & 16 & 32 & 64 & 128 & 256 & 512 & 1024 & 2048 & 4096 \\ \hline
16 & 43.0 & 36.4 & 30.1 & 25.1 & 25.3 & 24.8 & 25.2 & 25.0 & 35.8 \\ \hline
32 & 24.0 & 24.2 & 17.9 & 17.1 & 17.0 & 16.6 & 16.8 & 17.0 & 17.2 \\ \hline
64 & 18.2 & 15.3 & 13.0 & 13.5 & 13.4 & 13.2 & 12.9 & 13.1 & 18.0 \\ \hline
128 & 14.6 & 11.7 & 11.6 & 11.9 & 11.7 & 11.6 & 11.4 & 11.4 & 11.6 \\ \hline
256 & 12.5 & 11.5 & 11.4 & 11.4 & 10.8 & 10.6 & 10.6 & 10.7 & 11.9 \\ \hline
512 & 12.2 & 11.3 & 11.2 & 11.0 & 10.8 & 10.6 & 11.5 & 11.0 & 14.2 \\ \hline
1024 & 12.0 & 11.9 & 10.9 & 10.7 & 10.5 & 10.2 & 10.6 & 13.3 & 15.4 \\ \hline
2048 & 12.6 & 11.5 & 10.9 & 10.6 & 10.4 & 10.5 & 13.1 & 15.3 & 15.3 \\ \hline
4096 & 13.1 & 13.0 & 12.3 & 12.1 & 12.0 & 14.4 & 16.3 & 16.8 & 16.5 \\ \hline
\end{tabular}
}
\end{table}


This approach is justified by an experiment we conducted. In Table~\ref{tab:ab_matmul_param}, \textcolor{black}{for two machines} we record the performance of multiplying two matrices of dimensions $4096 \times 4096$ for each combination of $k_c$ and $n_c$ where both $k_c$ and $n_c$ are a power of 2 between 16 and 4096. \textcolor{black}{The first machine (top table) uses 512-bit SIMD registers and has packing disabled, and the second machine (bottom table) uses 256-bit SIMD registers and has packing enabled. For the first machine} we get the best performance when $k_c = 512$ and $n_c = 32$. We can find these parameters by first fixing $n_c$ to be 16 and finding the value of $k_c$ that yields the best performance, which is 512. From there we fix $k_c$ to be 512 and experiment with different values of $n_c$ to find the value that yields the best performance, and would see that choosing $n_c = 32$ yields the best performance compared to other choices for $n_c$. We would thus find a configuration of $k_c = 512, n_c=32$ which are the optimal parameter values for $k_c$ and $n_c$. In the following sections, we go into detail on how \mmlt{} finds suitable values for $k_c$ and $n_c$ at runtime.

\textcolor{black}{For the second machine, we can use the same strategy to get optimal parameter values $k_c = 1024, n_c = 512$. Compared to the first machine, slower runtimes are due to smaller SIMD register size, while larger optimal tiling parameter values are due to enhanced locality from packing.}

\subsection{Finding $k_c$}

\begin{figure}
    \centering
    \includegraphics[scale=0.165]{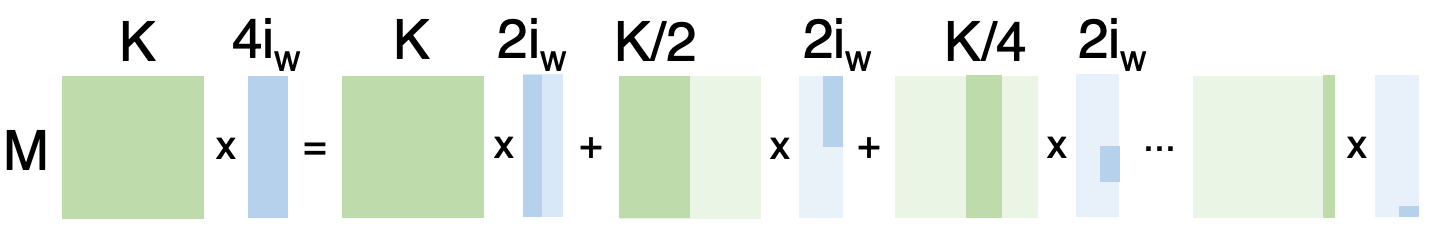}
    \caption{Subtasks used to test for $k_c$}
    \label{fig:kc_test}
\end{figure}

Given a matrix multiplication-like task that operates on two matrices of size $M \times K$ and $K \times N$, and when \mmlt{} generates an inner kernel of size $i_h \times i_w$, upon execution the task is partitioned into subtasks so that values of $k_c$ where $k_c = K,  \lceil K / 2 \rceil,  \lceil K / 4 \rceil,  \lceil K / 8 \rceil ...$ and $k_c \geq 16$ are tested while computing $M \times 4i_w$ results of the result matrix, as shown in Figure~\ref{fig:kc_test}. The last task in the figure exists to execute the remainder of the task and is not used to test a value of $k_c$. A subtask that operates on two matrices of size $M \times k$ and $k \times 2i_w$ tests when $k_c = k$. Each subtask is executed and a score of (execution time / k) is given to each subtask. The value of $k_c$ tested by the subtask with the lowest score is deemed the best value of $k_c$ as it means that using that particular value for $k_c$ resulted in the most efficient task execution.


\subsection{Finding $n_c$}

\begin{figure}
    \centering
    \includegraphics[scale=0.165]{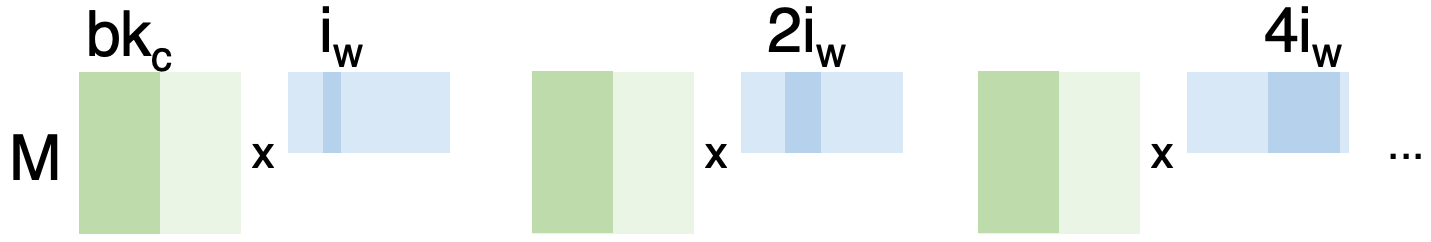}
    \caption{Subtasks used to test for $n_c$}
    \label{fig:nc_test}
\end{figure}

Let's denote the best value of $k_c$ found in the previous step as $bk_c$. To test for various values of $n_c$, we execute subtasks that operate on matrix pairs that have size ($M \times bk_c, bk_c \times n$) where $n$ is the value of $n_c$ that the subtask tests. To ensure correctness of results, we choose subtasks whose computations do not overlap with previously chosen subtasks. We test values of $n_c$ where $n_c = i_w, 2i_w, 4i_w ...$ and so on, as shown in Figure~\ref{fig:nc_test}. We start from $n_c = i_w$ and double the value of $n_c$ on each subsequent subtask. Similarly to the previous step, each time we complete a subtask we give that subtask a score of (execution time / n). If the score of a subtask is larger than the score of the previous subtask, we stop creating and executing subtasks and deem the value of $n_c$ of the previous subtask as the best value of $n_c$. This is because if the score of a subtask increases, it means that the tested value of $n_c$ is large enough to overflow the L2 cache which results in inefficient execution. Since choosing a larger value of $n_c$ will also result in cache overflow, we stop testing more values of $n_c$.

\mmlt{} then executes the remainder of the matrix multiplication-like task using the best values of $k_c$ and $n_c$ found in the previous steps.

\section{Implementation}\label{s:implementation}

We use the Truffle language implementation framework to build \mmlt{}. Truffle is an open source library for building program language implementations as interpreters for self-modifying Abstract Syntax Trees. It allows for relative ease of developing programming languages as the programmer only needs to implement nodes used for the AST of the programming language, but also provides competitive performance as well because compilation of the ASTs into fast machine code is supported by the Graal compiler.

In this paper we target declarative loops that are used within a JavaScript program. However, note that the techniques introduced in this paper are not limited to JavaScript, and could be used for any imperative programming language. \mmlt{} does not assume that consecutive rows in a matrix are stored contiguously, something that might true in languages like C but is not guaranteed in languages like Java.

\subsection{Preprocessing}

Upon execution of the user's JavaScript program, a custom script first rewrites the program source code to use the Polyglot API, an API that allows different languages implemented with Truffle to interoperate with each other ~\cite{polyglot}. In our implementation, we use Polyglot to access variables in JavaScript, and make the following changes to the source code: (1) The declarative loop itself is transformed into a string; (2) Values of all variables that are used in the declarative loop, but defined outside of the declarative loop, are stored in a dictionary; (3) This dictionary and the declarative loop string are passed to the code generation framework via the Polyglot API. 

To generate parameterized code for matrix multiplication tasks, the declarative loop string is parsed using a custom parser built with ANTLR~\cite{antlr} and the loop is checked to see whether it is a matrix multiplication-like task based on the conditions explained in Section~\ref{s:matmul task conditions}. Once \mmlt{} recognizes that a declarative loop is a matrix multiplication-like task, to generate code for the inner kernel the computation statement in the declarative loop is first parsed into a custom AST (not into a Truffle AST) and is then transformed into a DAG by performing common subexpression elimination. The DAG is then traversed using the \textit{labelfs} method introduced in ~\cite{kebler91arandomized} and the pseudo instructions needed to compute a single subresult are saved as an array of strings. How many extra registers are needed is also saved by the framework. The appropriate size of the inner kernel is then calculated using the method shown in Section~\ref{s:choosing kernel size}. Information about the inner kernel, i.e. instructions for calculating a subresult, kernel size, and matrix memory locations are then passed to the Graal compiler so that machine code for the inner kernel can be generated. However, due to the nature of the API provided by Graal only primitive data types can be communicated between the Truffle framework and the Graal compiler, so the instructions for calculating a subresult that were stored as strings are encoded into binary strings and then transformed into long integers and are then passed to the Graal compiler.

In the Graal compiler, we implement a function that accepts information about the inner kernel and generates assembly instructions for the actual inner kernel. The function generates instructions corresponding to a loop where execution subresults are accumulated in intermediate registers before being written to the result matrix after exiting the loop. The instructions for calculating a single subresult are generated by decoding the instructions that were encoded as long integers, and the same instructions are used for every subresult that needs to be calculated. How many subresults are calculated at a time is determined by the kernel size passed to the function. This function that generates inner kernel code is called when we pass information about the inner kernel to the Graal compiler from the Truffle framework.

To generate code for the parameterized nested loop that contains the inner kernel function, we implement a set of Truffle AST nodes, including value nodes (e.g. constants), arithmetic nodes (e.g. Addition) and condition nodes (e.g. LessThan). We also implement nodes corresponding to control flow, such as while nodes. We then create a Truffle AST corresponding to the nested loop that is suited for matrix multiplication-like tasks, with the inner kernel function being called at the innermost level of the loop.

The created Truffle AST is then compiled into fast machine code by the Graal compiler. \mmlt{} then executes this code using the adaptive strategy discussed in Section~\ref{adaptive_param_strategy}. Different portions of the task are executed by setting variables corresponding to the loop parameters (e.g. the start and end values of each nest of the loop) before executing the code, which allows the adaptive execution strategy to be possible.

\section{Experimental Evaluation}\label{s:experiment results}

We explain several experiments conducted on \mmlt{} in this section. All experiments were conducted on a Linux server with a 2.4GHz Xeon Gold 6312U processor. All experiments were executed with a single thread with hugepages on, and all data was stored as arrays in memory. We report the mean of 10 runs for each execution, excluding runs that take more than 6 hours to complete. In that case, we report results for a single run.

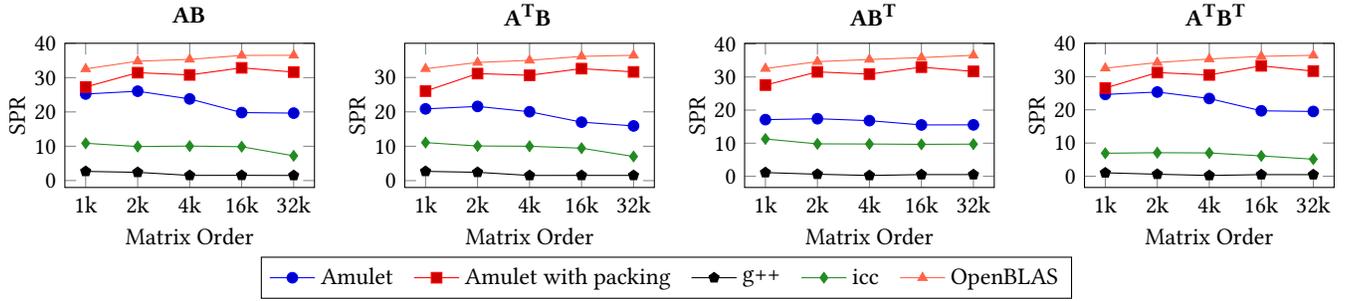
\begin{figure*}
\begin{tikzpicture}
    \definecolor{orangeclr}{RGB}{255,99,71}
    \definecolor{greenclr}{RGB}{34,139,34}
    \node[name=AB] at (1.65cm,2.3cm) {\textbf{AB}};
    \node[name=AtB] at (6.15cm,2.3cm) {\textbf{A\textsuperscript{T}B}};
    \node[name=ABt] at (10.75cm,2.3cm) {\textbf{AB\textsuperscript{T}}};
    \node[name=AtBt] at (15.3cm,2.3cm){\textbf{A\textsuperscript{T}B\textsuperscript{T}}};
	\begin{axis}[
	    name=ab,
		height=3.5cm,
		width=4.9cm,
		xticklabels={1k, 2k, 4k, 16k, 32k},
		xtick={1,2,3,4,5},
		xlabel={Matrix Order},
		xlabel style={ yshift=1mm, },
		ylabel={SPR},
		ylabel style={ yshift=-6mm, },
		legend style={at={(2.41,-0.48)},
      anchor=north,legend columns=-1,/tikz/every even column/.append style={column sep=0.2cm}},
	]
	\addplot table [x=size, y=mmlt, col sep=comma] {AB_matmul_data.csv};
	\addplot table [x=size, y=packed, col sep=comma] {AB_matmul_data.csv};
	\addplot [mark=pentagon*] table [x=size, y=g++, col sep=comma] {AB_matmul_data.csv};
	\addplot [greenclr, mark=diamond*] table [x=size, y=icc, col sep=comma] {AB_matmul_data.csv};
	\addplot [orangeclr, mark=triangle*] table [x=size, y=openblas, col sep=comma] {AB_matmul_data.csv};
	\legend{\mmlt{},\mmlt{} with packing,g++,icc,OpenBLAS}
	\end{axis}
	
	\begin{axis}[
	    name=atb,
	    at={(ab.south east)},
	    xshift=1.2cm,
		height=3.5cm,
		width=4.9cm,
		xticklabels={1k, 2k, 4k, 16k, 32k},
		xtick={1,2,3,4,5},
		xlabel={Matrix Order},
		xlabel style={ yshift=1mm, },
		ylabel={SPR},
		ylabel style={ yshift=-6mm, },
		legend style={at={(0.5,-0.23)},
      anchor=north,legend columns=-1,/tikz/every even column/.append style={column sep=0.2cm}},
	]
	\addplot table [x=size, y=mmlt, col sep=comma] {AtB_matmul_data.csv};
	\addplot table [x=size, y=packed, col sep=comma] {AtB_matmul_data.csv};
	\addplot [mark=pentagon*] table [x=size, y=g++, col sep=comma] {AtB_matmul_data.csv};
	\addplot [greenclr, mark=diamond*] table [x=size, y=icc, col sep=comma] {AtB_matmul_data.csv};
	\addplot [orangeclr, mark=triangle*] table [x=size, y=openblas, col sep=comma] {AtB_matmul_data.csv};
	\end{axis}

	\begin{axis}[
	    name=abt,
	    at={(atb.south east)},
	    xshift=1.2cm,
		height=3.5cm,
		width=4.9cm,
		xticklabels={1k, 2k, 4k, 16k, 32k},
		xtick={1,2,3,4,5},
		xlabel={Matrix Order},
		xlabel style={ yshift=1mm, },
		ylabel={SPR},
		ylabel style={ yshift=-6mm, },
		legend style={at={(0.5,-0.23)},
      anchor=north,legend columns=-1,/tikz/every even column/.append style={column sep=0.2cm}},
	]
	\addplot table [x=size, y=mmlt, col sep=comma] {ABt_matmul_data.csv};
	\addplot table [x=size, y=packed, col sep=comma] {ABt_matmul_data.csv};
	\addplot [mark=pentagon*] table [x=size, y=g++, col sep=comma] {ABt_matmul_data.csv};
	\addplot [greenclr, mark=diamond*] table [x=size, y=icc, col sep=comma] {ABt_matmul_data.csv};
	\addplot [orangeclr, mark=triangle*] table [x=size, y=openblas, col sep=comma] {ABt_matmul_data.csv};
	\end{axis}
	
	\begin{axis}[
	    name=atbt,
	    at={(abt.south east)},
	    xshift=1.2cm,
		height=3.5cm,
		width=4.9cm,
		xticklabels={1k, 2k, 4k, 16k, 32k},
		xtick={1,2,3,4,5},
		xlabel={Matrix Order},
		xlabel style={ yshift=1mm, },
		ylabel={SPR},
		ylabel style={ yshift=-6mm, },
		legend style={at={(0.5,-0.23)},
      anchor=north,legend columns=-1,/tikz/every even column/.append style={column sep=0.2cm}},
	]
	\addplot table [x=size, y=mmlt, col sep=comma] {AtBt_matmul_data.csv};
	\addplot table [x=size, y=packed, col sep=comma] {AtBt_matmul_data.csv};
	\addplot [mark=pentagon*] table [x=size, y=g++, col sep=comma] {AtBt_matmul_data.csv};
	\addplot [greenclr, mark=diamond*] table [x=size, y=icc, col sep=comma] {AtBt_matmul_data.csv};
	\addplot [orangeclr, mark=triangle*] table [x=size, y=openblas, col sep=comma] {AtBt_matmul_data.csv};
	\end{axis}
	
\end{tikzpicture}
\caption{Matrix Multiplication Performance (Higher is better)}
\label{fig:matmul performance linechart}
\vspace*{-2mm}
\end{figure*}

\subsection{Matrix Multiplication Performance}

We ran experiments to see how well \mmlt{} compares to conventional ways of doing matrix multiplication. We use the following baselines: A simple triple nested loop written in C compiled with g++, the same loop compiled with the Intel C++ Compiler (icc) ~\cite{intelicc}, (for both compilers, we use the -O3 option as it produced the fastest code out of the various flags/options we tried) and OpenBLAS, a popular library for performing linear algebra. For the triple nested loop written in C, we report experiment results using the loop order (out of 6 possible orders) that resulted in the shortest execution time. We also did experiments using Intel's next generation compiler (icx) ~\cite{intelicx} but omitted the result as we found that icc performed better. We also performed experiments using the Intel Math Kernel Library (MKL) ~\cite{mkl} and NumPy ~\cite{NumPy} but similarly omitted the results as we found that OpenBLAS always performed similarly or slightly better. We also experimented using Postgres where we store the matrices as relations and use a JOIN query to do matrix multiplication. This is analogous to the situation where the user utilizes a database system to perform data analysis tasks. However we have also omitted this result as even when operating on relatively small matrices of dimensions 1024x1024 the query took more than 300 seconds to complete, which is more than $7000 \times$ slower than \mmlt{}.
The slow performance of Postgres is due to the creation of intermediate data structures and slow access patterns of joins and aggregations.

Let the dimensions of the two matrices to be multiplied be expressed as $M \times K$ and $K \times N$. We experimented by multiplying square matrices (i.e. $M = K = N$) of order 1024, 2048, 4096, 16384, and 32768 (referred to as 1k, 2k, 4k, 16k, 32k). (We also performed experiments where the matrices to be multiplied are not square in Section~\ref{non-square}.) For each order we tried all four combinations of whether the matrices to be multiplied were row-major or column-major (for example, $AB^T$ means that the first matrix is row-major and the second matrix is column-major). We measure performance in terms of \textbf{Scaled Processing Rate (SPR)}, where given a matrix multiplication-like task that operates on matrices of dimension $M \times K$ and $K \times N$, and its execution time $T$, SPR is calculated as $\mathbf{(MKN)/(10^9T)}$. For example, if multiplying two 1000x1000 matrices takes 0.1 seconds, the Scaled Processing Rate would be 10. Since the methods of execution we experiment with do matrix multiplication in $O(MKN)$ time, using Scaled Processing Rate as a performance metric allows us to (1) evaluate whether execution time scales appropriately with matrix size and (2) compare execution time across different execution methods.

Results are shown in Figure~\ref{fig:matmul performance linechart}. We also include performance results of \mmlt{} enabled with the packing technique explained in section~\ref{libraryadditionaltechniques}. For each method of execution, we plot a linechart where the x axis value is the order of the matrices to be multiplied, and the y axis value is the performance in terms of Scaled Processing Rate. \mmlt{} without packing is on average $2.39 \times$ faster than icc, and $26.37 \times$ faster than g++. We observed that both g++ and icc generate vectorized code, and icc also does automatic loop tiling with static tile sizes, which explains icc's low execution time compared to g++.
Nevertheless, loop tiling alone does not match the full performance impact of \mmlt{}: \mmlt{} also benefits from adaptive parameter selection, register optimization, and use of the Goto et al. algorithm~\cite{GotoMat08}.

For smaller matrices, \mmlt{} without packing has reasonable performance (approximately 30\% to 50\% slower) compared to the hand-tuned OpenBLAS library. As the matrices grow larger, OpenBLAS becomes comparatively faster than \mmlt{} as the overhead of matrix packing becomes smaller compared to the overall execution time, while \mmlt{} suffers more TLB misses as the memory footprint grows larger. With packing enabled, these TLB misses decrease, and the performance gap between \mmlt{} and OpenBLAS stays relatively consistent, with \mmlt{} being approximately 10\% to 20\% slower than OpenBLAS regardless of matrix size. Profiling results for matrix multiplication on matrices of order 2048 are given in Table~\ref{tab:performance stats} for reference. The remaining performance gap is likely due to OpenBLAS also hand-tuning parameters for prefetching, as well as interleaving packing and execution to further increase cache locality, which due to its complex implementation, is not currently supported by \mmlt{}. \textcolor{black}{Note that OpenBLAS and other matrix multiplication libraries cannot support matrix multiplication-like tasks other than matrix multiplication. With a simple engineering change we can make \mmlt{} use the OpenBLAS API when matrix multiplication is detected. The purpose of running experiments on matrix multiplication is to give confidence that our approach has low performance overhead.}

\begin{table}[]
\centering
\caption{Profiling results for \mmlt{} and OpenBLAS}
\label{tab:performance stats}
\setlength\tabcolsep{2.5pt}
\def\arraystretch{1.1}%
\begin{tabular}{|c|c|c|c|}
\hline
{\backslashbox{Event}{Method}} & L1 cache misses & LLC cache misses & dTLB misses\\ \thickhline
\mmlt{} & 366232k & 2467k & 3927k \\ \thickhline
\mmlt{} w/ packing & 243248k & 1822k & 841k \\ \thickhline
OpenBLAS & 236718k & 902k & 517k \\ \thickhline
\end{tabular}
\vspace{-1mm}
\end{table}

\subsection{Effectiveness of Adaptive Execution Strategy}

\begin{table}[]
\centering
\caption{Effectiveness of Adaptive Execution}
\label{tab:adaptive execution}
\def\arraystretch{1.1}%
\begin{threeparttable}
\begin{tabular}{|P{1cm}|c|c|c|c|}
\hline
\multicolumn{2}{|c|}{\backslashbox{Form/Method}{Matrix Order}} & 1k & 2k & 4k\\ \thickhline
\multirow{2}{*}{$AB$} & Manual & 0.041 & 0.327 & 2.781 \\ \cline{2-5}
& Adaptive & 0.042 & 0.330 & 2.889 \\ \thickhline
\multirow{2}{*}{$A^TB$} &Manual & 0.049 & 0.389 & 3.213 \\ \cline{2-5}
& Adaptive & 0.051 & 0.398 & 3.423 \\ \thickhline
\multirow{2}{*}{$AB^T$} & Manual & 0.060 & 0.481 & 4.002 \\ \cline{2-5}
& Adaptive & 0.063 & 0.494 & 4.089 \\ \thickhline
\multirow{2}{*}{$A^TB^T$} & Manual & 0.042 & 0.328 & 2.902 \\ \cline{2-5}
& Adaptive & 0.044 & 0.339 & 2.937 \\ \thickhline
\end{tabular}
\begin{tablenotes}[para,flushright]
     Unit: Seconds
\end{tablenotes}
\end{threeparttable}
\end{table}

To see how effective our adaptive strategy is for finding the right parameters for code generated for matrix multiplication-like tasks, we manually tested many parameters for matrix multiplication (as shown in Section~\ref{adaptive_param_strategy}) for square matrices of order 1k, 2k, and 4k and saw how the best performance we found compares with the adaptive strategy. Results are shown in Table~\ref{tab:adaptive execution}. `Manual' represents the best performance found through manual testing, and `adaptive' represents the performance of \mmlt{}.

We can see that for all configurations of how the matrices are stored, the adaptive strategy works well as it yields execution time within 7\% of the best execution time found through manual testing. 



\subsection{Performance on general code}\label{s:general code results}

\begin{figure}
\begin{Verbatim}[commandchars=\\\{\}, frame=single, fontsize=\small]
Query 1
where(i in [0..M] and j in [0..N] and k in [0..K]) \{
  R[i][j] += \textcolor{red}{A[i][k]}*\textcolor{blue}{B[k][j]}-
  (\textcolor{red}{A[i][k]}*\textcolor{blue}{B[k][j]}>\textcolor{green!45!black}{thres[j]})*\textcolor{red}{A[i][k]}*\textcolor{blue}{B[k][j]}*dis[j];
\}

Query 2
where(i in [0..M] and j in [0..N] and k in [0..K]) \{
  R[i][j] += \textcolor{red}{A[i][k]}*\textcolor{blue}{B[k][j]}+
  (\textcolor{red}{A[i][k]}*\textcolor{blue}{B[k][j]}>\textcolor{green!45!black}{thres[j]})*(\textcolor{red}{A[i][k]}*\textcolor{blue}{B[k][j]}-\textcolor{green!45!black}{thres[j]});
\}

Query 3
where(i in [0..M] and j in [0..N] and k in [0..K]) \{
  R[i][j] += (\textcolor{red}{A[i][k]}*\textcolor{blue}{B[k][j]}>\textcolor{green!45!black}{100});
\}

\textbullet When A is transposed, \textcolor{red}{A[i][k]} is replaced with \textcolor{red}{A[k][i]}
\textbullet When B is transposed, \textcolor{blue}{B[k][j]} is replaced with \textcolor{blue}{B[j][k]}
\textbullet Variable in \textcolor{green!45!black}{green} can be replaced with 
  \textcolor{green!45!black}{100}, \textcolor{green!45!black}{thres[i]}, \textcolor{green!45!black}{thres[j]}, \textcolor{green!45!black}{thres[i][j]}.
\end{Verbatim}
\caption{Test Queries}
\vspace*{-1mm}
\label{fig:test_queries}
\end{figure}

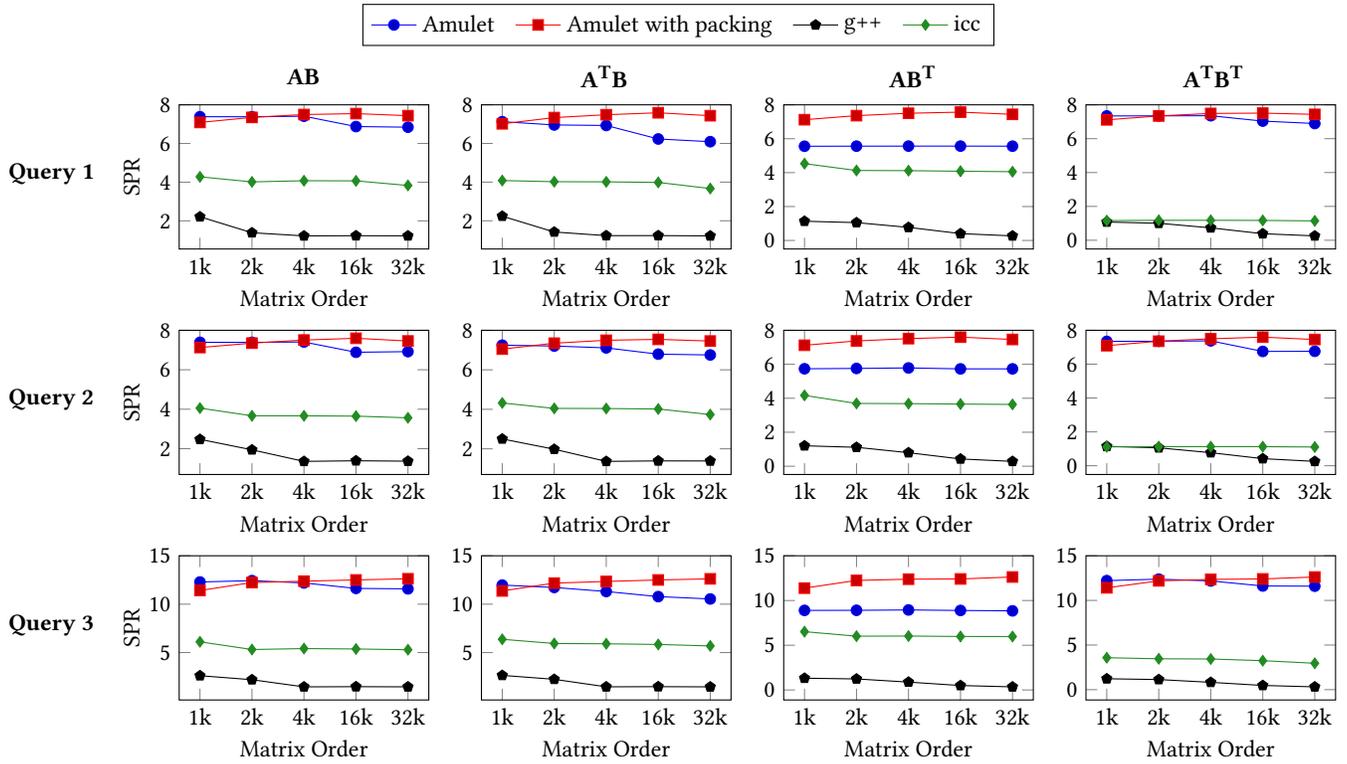
\begin{figure*}
\begin{tikzpicture}
    \definecolor{greenclr}{RGB}{34,139,34}
    \node[name=query] at (0cm,1cm) {\textbf{Query 1}};
    \node[name=query] at (0cm,-2cm) {\textbf{Query 2}};
    \node[name=query] at (0cm,-5cm) {\textbf{Query 3}};
    \node[name=AB] at (3.35cm,2.3cm) {\textbf{AB}};
    \node[name=AtB] at (7.35,2.3cm) {\textbf{A\textsuperscript{T}B}};
    \node[name=ABt] at (11.45,2.3cm) {\textbf{AB\textsuperscript{T}}};
    \node[name=AtBt] at (15.45cm,2.3cm){\textbf{A\textsuperscript{T}B\textsuperscript{T}}};
	\begin{axis}[
	    name=abq1,
	    at={(1.7cm, 0cm)},
		height=3.5cm,
		width=4.9cm,
		xticklabels={1k, 2k, 4k, 16k, 32k},
		xtick={1,2,3,4,5},
		xlabel={Matrix Order},
		xlabel style={ yshift=1mm, },
		ylabel={SPR},
		ymax=8,
		ylabel style={ yshift=-6mm, },
		legend style={at={(2.00,1.7)},
      anchor=north,legend columns=-1,/tikz/every even column/.append style={column sep=0.2cm}},
	]
	\addplot table [x=size, y=mmlt, col sep=comma] {AB_q1_data.csv};
	\addplot table [x=size, y=packed, col sep=comma] {AB_q1_data.csv};
	\addplot [mark=pentagon*] table [x=size, y=g++, col sep=comma] {AB_q1_data.csv};
	\addplot [greenclr, mark=diamond*] table [x=size, y=icc, col sep=comma] {AB_q1_data.csv};
	\legend{\mmlt{},\mmlt{} with packing,g++,icc}
	\end{axis}
	
	\begin{axis}[
	    name=atbq1,
	    at={($(abq1.south east)+(-0.5cm,0cm)$)},
	    xshift=1.2cm,
		height=3.5cm,
		width=4.9cm,
		xticklabels={1k, 2k, 4k, 16k, 32k},
		xtick={1,2,3,4,5},
		xlabel={Matrix Order},
		xlabel style={ yshift=1mm, },
		ymax=8,
		ylabel style={ yshift=-6mm, },
		legend style={at={(0.5,-0.23)},
      anchor=north,legend columns=-1,/tikz/every even column/.append style={column sep=0.2cm}},
	]
	\addplot table [x=size, y=mmlt, col sep=comma] {AtB_q1_data.csv};
	\addplot table [x=size, y=packed, col sep=comma] {AtB_q1_data.csv};
	\addplot [mark=pentagon*] table [x=size, y=g++, col sep=comma] {AtB_q1_data.csv};
	\addplot [greenclr, mark=diamond*] table [x=size, y=icc, col sep=comma] {AtB_q1_data.csv};
	\end{axis}

	\begin{axis}[
	    name=abtq1,
	    at={($(atbq1.south east)+(-0.5cm,0cm)$)},
	    xshift=1.2cm,
		height=3.5cm,
		width=4.9cm,
		xticklabels={1k, 2k, 4k, 16k, 32k},
		xtick={1,2,3,4,5},
		xlabel={Matrix Order},
		xlabel style={ yshift=1mm, },
		ymax=8,
		ylabel style={ yshift=-6mm, },
		legend style={at={(0.5,-0.23)},
      anchor=north,legend columns=-1,/tikz/every even column/.append style={column sep=0.2cm}},
	]
	\addplot table [x=size, y=mmlt, col sep=comma] {ABt_q1_data.csv};
	\addplot table [x=size, y=packed, col sep=comma] {ABt_q1_data.csv};
	\addplot [mark=pentagon*] table [x=size, y=g++, col sep=comma] {ABt_q1_data.csv};
	\addplot [greenclr, mark=diamond*] table [x=size, y=icc, col sep=comma] {ABt_q1_data.csv};
	\end{axis}
	
	\begin{axis}[
	    name=atbtq1,
	    at={($(abtq1.south east)+(-0.5cm,0cm)$)},
	    xshift=1.2cm,
		height=3.5cm,
		width=4.9cm,
		xticklabels={1k, 2k, 4k, 16k, 32k},
		xtick={1,2,3,4,5},
		xlabel={Matrix Order},
		xlabel style={ yshift=1mm, },
		ymax=8,
		ylabel style={ yshift=-6mm, },
		legend style={at={(0.5,-0.23)},
      anchor=north,legend columns=-1,/tikz/every even column/.append style={column sep=0.2cm}},
	]
	\addplot table [x=size, y=mmlt, col sep=comma] {AtBt_q1_data.csv};
	\addplot table [x=size, y=packed, col sep=comma] {AtBt_q1_data.csv};
	\addplot [mark=pentagon*] table [x=size, y=g++, col sep=comma] {AtBt_q1_data.csv};
	\addplot [greenclr, mark=diamond*] table [x=size, y=icc, col sep=comma] {AtBt_q1_data.csv};
	\end{axis}

	\begin{axis}[
	    name=abq2,
	    at={($(abq1.south west)+(0cm,-3cm)$)},
		height=3.5cm,
		width=4.9cm,
		xticklabels={1k, 2k, 4k, 16k, 32k},
		xtick={1,2,3,4,5},
		xlabel={Matrix Order},
		xlabel style={ yshift=1mm, },
		ymax=8,
		ylabel={SPR},
		ylabel style={ yshift=-6mm, },
		legend style={at={(2.41,-0.33)},
      anchor=north,legend columns=-1,/tikz/every even column/.append style={column sep=0.2cm}},
	]
	\addplot table [x=size, y=mmlt, col sep=comma] {AB_q2_data.csv};
	\addplot table [x=size, y=packed, col sep=comma] {AB_q2_data.csv};
	\addplot [mark=pentagon*] table [x=size, y=g++, col sep=comma] {AB_q2_data.csv};
	\addplot [greenclr, mark=diamond*] table [x=size, y=icc, col sep=comma] {AB_q2_data.csv};
	\end{axis}
	
	\begin{axis}[
	    name=atbq2,
	    at={($(abq2.south east)+(-0.5cm,0cm)$)},
	    xshift=1.2cm,
		height=3.5cm,
		width=4.9cm,
		xticklabels={1k, 2k, 4k, 16k, 32k},
		xtick={1,2,3,4,5},
		xlabel={Matrix Order},
		xlabel style={ yshift=1mm, },
		ymax=8,
		ylabel style={ yshift=-6mm, },
		legend style={at={(0.5,-0.23)},
      anchor=north,legend columns=-1,/tikz/every even column/.append style={column sep=0.2cm}},
	]
	\addplot table [x=size, y=mmlt, col sep=comma] {AtB_q2_data.csv};
	\addplot table [x=size, y=packed, col sep=comma] {AtB_q2_data.csv};
	\addplot [mark=pentagon*] table [x=size, y=g++, col sep=comma] {AtB_q2_data.csv};
	\addplot [greenclr, mark=diamond*] table [x=size, y=icc, col sep=comma] {AtB_q2_data.csv};
	\end{axis}

	\begin{axis}[
	    name=abtq2,
	    at={($(atbq2.south east)+(-0.5cm,0cm)$)},
	    xshift=1.2cm,
		height=3.5cm,
		width=4.9cm,
		xticklabels={1k, 2k, 4k, 16k, 32k},
		xtick={1,2,3,4,5},
		xlabel={Matrix Order},
		xlabel style={ yshift=1mm, },
		ymax=8,
		ylabel style={ yshift=-6mm, },
		legend style={at={(0.5,-0.23)},
      anchor=north,legend columns=-1,/tikz/every even column/.append style={column sep=0.2cm}},
	]
	\addplot table [x=size, y=mmlt, col sep=comma] {ABt_q2_data.csv};
	\addplot table [x=size, y=packed, col sep=comma] {ABt_q2_data.csv};
	\addplot [mark=pentagon*] table [x=size, y=g++, col sep=comma] {ABt_q2_data.csv};
	\addplot [greenclr, mark=diamond*] table [x=size, y=icc, col sep=comma] {ABt_q2_data.csv};
	\end{axis}
	
	\begin{axis}[
	    name=atbtq2,
	    at={($(abtq2.south east)+(-0.5cm,0cm)$)},
	    xshift=1.2cm,
		height=3.5cm,
		width=4.9cm,
		xticklabels={1k, 2k, 4k, 16k, 32k},
		xtick={1,2,3,4,5},
		xlabel={Matrix Order},
		xlabel style={ yshift=1mm, },
		ymax=8,
		ylabel style={ yshift=-6mm, },
		legend style={at={(0.5,-0.23)},
      anchor=north,legend columns=-1,/tikz/every even column/.append style={column sep=0.2cm}},
	]
	\addplot table [x=size, y=mmlt, col sep=comma] {AtBt_q2_data.csv};
	\addplot table [x=size, y=packed, col sep=comma] {AtBt_q2_data.csv};
	\addplot [mark=pentagon*] table [x=size, y=g++, col sep=comma] {AtBt_q2_data.csv};
	\addplot [greenclr, mark=diamond*] table [x=size, y=icc, col sep=comma] {AtBt_q2_data.csv};
	\end{axis}

	\begin{axis}[
	    name=abq3,
	    at={($(abq2.south west)+(0cm,-3cm)$)},
		height=3.5cm,
		width=4.9cm,
		xticklabels={1k, 2k, 4k, 16k, 32k},
		xtick={1,2,3,4,5},
		xlabel={Matrix Order},
		xlabel style={ yshift=1mm, },
		ymax=15,
		ylabel={SPR},
		ylabel style={ yshift=-6mm, },
		legend style={at={(2.41,-0.33)},
      anchor=north,legend columns=-1,/tikz/every even column/.append style={column sep=0.2cm}},
	]
	\addplot table [x=size, y=mmlt, col sep=comma] {AB_q3_data.csv};
	\addplot table [x=size, y=packed, col sep=comma] {AB_q3_data.csv};
	\addplot [mark=pentagon*] table [x=size, y=g++, col sep=comma] {AB_q3_data.csv};
	\addplot [greenclr, mark=diamond*] table [x=size, y=icc, col sep=comma] {AB_q3_data.csv};
	\end{axis}
	
	\begin{axis}[
	    name=atbq3,
	    at={($(abq3.south east)+(-0.5cm,0cm)$)},
	    xshift=1.2cm,
		height=3.5cm,
		width=4.9cm,
		xticklabels={1k, 2k, 4k, 16k, 32k},
		xtick={1,2,3,4,5},
		xlabel={Matrix Order},
		xlabel style={ yshift=1mm, },
		ymax=15,
		ylabel style={ yshift=-6mm, },
		legend style={at={(0.5,-0.23)},
      anchor=north,legend columns=-1,/tikz/every even column/.append style={column sep=0.2cm}},
	]
	\addplot table [x=size, y=mmlt, col sep=comma] {AtB_q3_data.csv};
	\addplot table [x=size, y=packed, col sep=comma] {AtB_q3_data.csv};
	\addplot [mark=pentagon*] table [x=size, y=g++, col sep=comma] {AtB_q3_data.csv};
	\addplot [greenclr, mark=diamond*] table [x=size, y=icc, col sep=comma] {AtB_q3_data.csv};
	\end{axis}

	\begin{axis}[
	    name=abtq3,
	    at={($(atbq3.south east)+(-0.5cm,0cm)$)},
	    xshift=1.2cm,
		height=3.5cm,
		width=4.9cm,
		xticklabels={1k, 2k, 4k, 16k, 32k},
		xtick={1,2,3,4,5},
		xlabel={Matrix Order},
		xlabel style={ yshift=1mm, },
		ymax=15,
		ylabel style={ yshift=-6mm, },
		legend style={at={(0.5,-0.23)},
      anchor=north,legend columns=-1,/tikz/every even column/.append style={column sep=0.2cm}},
	]
	\addplot table [x=size, y=mmlt, col sep=comma] {ABt_q3_data.csv};
	\addplot table [x=size, y=packed, col sep=comma] {ABt_q3_data.csv};
	\addplot [mark=pentagon*] table [x=size, y=g++, col sep=comma] {ABt_q3_data.csv};
	\addplot [greenclr, mark=diamond*] table [x=size, y=icc, col sep=comma] {ABt_q3_data.csv};
	\end{axis}
	
	\begin{axis}[
	    name=atbtq3,
	    at={($(abtq3.south east)+(-0.5cm,0cm)$)},
	    xshift=1.2cm,
		height=3.5cm,
		width=4.9cm,
		xticklabels={1k, 2k, 4k, 16k, 32k},
		xtick={1,2,3,4,5},
		xlabel={Matrix Order},
		xlabel style={ yshift=1mm, },
		ymax=15,
		ylabel style={ yshift=-6mm, },
		legend style={at={(0.5,-0.23)},
      anchor=north,legend columns=-1,/tikz/every even column/.append style={column sep=0.2cm}},
	]
	\addplot table [x=size, y=mmlt, col sep=comma] {AtBt_q3_data.csv};
	\addplot table [x=size, y=packed, col sep=comma] {AtBt_q3_data.csv};
	\addplot [mark=pentagon*] table [x=size, y=g++, col sep=comma] {AtBt_q3_data.csv};
	\addplot [greenclr, mark=diamond*] table [x=size, y=icc, col sep=comma] {AtBt_q3_data.csv};
	\end{axis}
	
\end{tikzpicture}
\vspace*{-5mm}
\caption{General Query Performance (Higher is better)}
\label{fig:general query performance linechart}
\end{figure*}

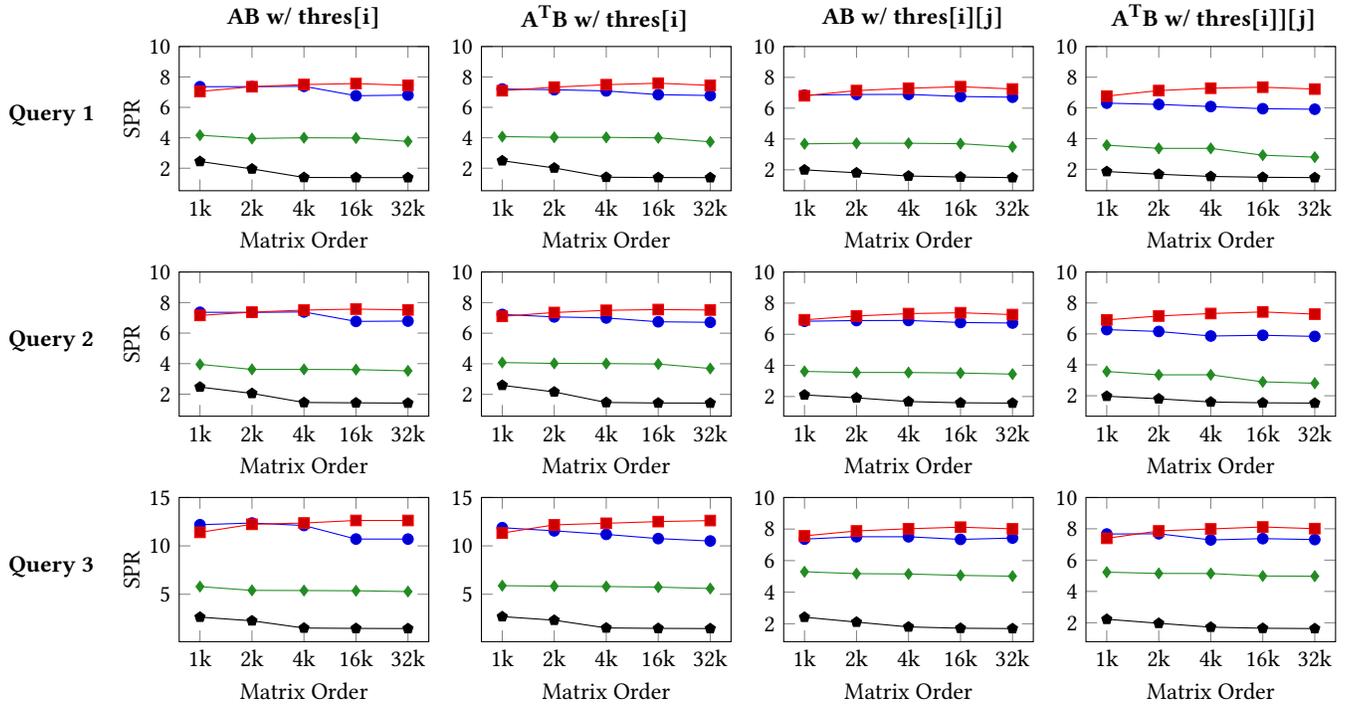
\begin{figure*}
\begin{tikzpicture}
    \definecolor{greenclr}{RGB}{34,139,34}
    \node[name=query] at (0cm,1cm) {\textbf{Query 1}};
    \node[name=query] at (0cm,-2cm) {\textbf{Query 2}};
    \node[name=query] at (0cm,-5cm) {\textbf{Query 3}};
    \node[name=ABthresi] at (3.35cm,2.3cm) {\textbf{AB w/ thres[i]}};
    \node[name=AtBthresi] at (7.35,2.3cm) {\textbf{A\textsuperscript{T}B w/ thres[i]}};
    \node[name=ABthresij] at (11.45cm,2.3cm) {\textbf{AB w/ thres[i][j]}};
    \node[name=AtBthresij] at (15.45,2.3cm) {\textbf{A\textsuperscript{T}B w/ thres[i]][j]}};
	\begin{axis}[
	    name=abq1,
	    at={(1.7cm, 0cm)},
		height=3.5cm,
		width=4.9cm,
		xticklabels={1k, 2k, 4k, 16k, 32k},
		xtick={1,2,3,4,5},
		xlabel={Matrix Order},
		xlabel style={ yshift=1mm, },
		ylabel={SPR},
		ymax=10,
		ylabel style={ yshift=-6mm, },
		legend style={at={(2.41,1.43)},
      anchor=north,legend columns=-1,/tikz/every even column/.append style={column sep=0.2cm}},
	]
	\addplot table [x=size, y=mmlt, col sep=comma] {ABi_q1_data.csv};
	\addplot table [x=size, y=packed, col sep=comma] {ABi_q1_data.csv};
	\addplot [mark=pentagon*] table [x=size, y=g++, col sep=comma] {ABi_q1_data.csv};
	\addplot [greenclr, mark=diamond*] table [x=size, y=icc, col sep=comma] {ABi_q1_data.csv};
	\end{axis}
	
	\begin{axis}[
	    name=atbq1,
	    at={($(abq1.south east)+(-0.5cm,0cm)$)},
	    xshift=1.2cm,
		height=3.5cm,
		width=4.9cm,
		xticklabels={1k, 2k, 4k, 16k, 32k},
		xtick={1,2,3,4,5},
		xlabel={Matrix Order},
		xlabel style={ yshift=1mm, },
		ymax=10,
		ylabel style={ yshift=-6mm, },
		legend style={at={(0.5,-0.23)},
      anchor=north,legend columns=-1,/tikz/every even column/.append style={column sep=0.2cm}},
	]
	\addplot table [x=size, y=mmlt, col sep=comma] {AtBi_q1_data.csv};
	\addplot table [x=size, y=packed, col sep=comma] {AtBi_q1_data.csv};
	\addplot [mark=pentagon*] table [x=size, y=g++, col sep=comma] {AtBi_q1_data.csv};
	\addplot [greenclr, mark=diamond*] table [x=size, y=icc, col sep=comma] {AtBi_q1_data.csv};
	\end{axis}

	\begin{axis}[
	    name=abtq1,
	    at={($(atbq1.south east)+(-0.5cm,0cm)$)},
	    xshift=1.2cm,
		height=3.5cm,
		width=4.9cm,
		xticklabels={1k, 2k, 4k, 16k, 32k},
		xtick={1,2,3,4,5},
		xlabel={Matrix Order},
		xlabel style={ yshift=1mm, },
		ymax=10,
		ylabel style={ yshift=-6mm, },
		legend style={at={(0.5,-0.23)},
      anchor=north,legend columns=-1,/tikz/every even column/.append style={column sep=0.2cm}},
	]
	\addplot table [x=size, y=mmlt, col sep=comma] {ABij_q1_data.csv};
	\addplot table [x=size, y=packed, col sep=comma] {ABij_q1_data.csv};
	\addplot [mark=pentagon*] table [x=size, y=g++, col sep=comma] {ABij_q1_data.csv};
	\addplot [greenclr, mark=diamond*] table [x=size, y=icc, col sep=comma] {ABij_q1_data.csv};
	\end{axis}
	
	\begin{axis}[
	    name=atbtq1,
	    at={($(abtq1.south east)+(-0.5cm,0cm)$)},
	    xshift=1.2cm,
		height=3.5cm,
		width=4.9cm,
		xticklabels={1k, 2k, 4k, 16k, 32k},
		xtick={1,2,3,4,5},
		xlabel={Matrix Order},
		xlabel style={ yshift=1mm, },
		ymax=10,
		ylabel style={ yshift=-6mm, },
		legend style={at={(0.5,-0.23)},
      anchor=north,legend columns=-1,/tikz/every even column/.append style={column sep=0.2cm}},
	]
	\addplot table [x=size, y=mmlt, col sep=comma] {AtBij_q1_data.csv};
	\addplot table [x=size, y=packed, col sep=comma] {AtBij_q1_data.csv};
	\addplot [mark=pentagon*] table [x=size, y=g++, col sep=comma] {AtBij_q1_data.csv};
	\addplot [greenclr, mark=diamond*] table [x=size, y=icc, col sep=comma] {AtBij_q1_data.csv};
	
	\end{axis}

	\begin{axis}[
	    name=abq2,
	    at={($(abq1.south west)+(0cm,-3cm)$)},
		height=3.5cm,
		width=4.9cm,
		xticklabels={1k, 2k, 4k, 16k, 32k},
		xtick={1,2,3,4,5},
		xlabel={Matrix Order},
		xlabel style={ yshift=1mm, },
		ymax=10,
		ylabel={SPR},
		ylabel style={ yshift=-6mm, },
		legend style={at={(2.41,-0.33)},
      anchor=north,legend columns=-1,/tikz/every even column/.append style={column sep=0.2cm}},
	]
	\addplot table [x=size, y=mmlt, col sep=comma] {ABi_q2_data.csv};
	\addplot table [x=size, y=packed, col sep=comma] {ABi_q2_data.csv};
	\addplot [mark=pentagon*] table [x=size, y=g++, col sep=comma] {ABi_q2_data.csv};
	\addplot [greenclr, mark=diamond*] table [x=size, y=icc, col sep=comma] {ABi_q2_data.csv};
	\end{axis}
	
	\begin{axis}[
	    name=atbq2,
	    at={($(abq2.south east)+(-0.5cm,0cm)$)},
	    xshift=1.2cm,
		height=3.5cm,
		width=4.9cm,
		xticklabels={1k, 2k, 4k, 16k, 32k},
		xtick={1,2,3,4,5},
		xlabel={Matrix Order},
		xlabel style={ yshift=1mm, },
		ymax=10,
		ylabel style={ yshift=-6mm, },
		legend style={at={(0.5,-0.23)},
      anchor=north,legend columns=-1,/tikz/every even column/.append style={column sep=0.2cm}},
	]
	\addplot table [x=size, y=mmlt, col sep=comma] {AtBi_q2_data.csv};
	\addplot table [x=size, y=packed, col sep=comma] {AtBi_q2_data.csv};
	\addplot [mark=pentagon*] table [x=size, y=g++, col sep=comma] {AtBi_q2_data.csv};
	\addplot [greenclr, mark=diamond*] table [x=size, y=icc, col sep=comma] {AtBi_q2_data.csv};
	\end{axis}

	\begin{axis}[
	    name=abtq2,
	    at={($(atbq2.south east)+(-0.5cm,0cm)$)},
	    xshift=1.2cm,
		height=3.5cm,
		width=4.9cm,
		xticklabels={1k, 2k, 4k, 16k, 32k},
		xtick={1,2,3,4,5},
		xlabel={Matrix Order},
		xlabel style={ yshift=1mm, },
		ymax=10,
		ylabel style={ yshift=-6mm, },
		legend style={at={(0.5,-0.23)},
      anchor=north,legend columns=-1,/tikz/every even column/.append style={column sep=0.2cm}},
	]
	\addplot table [x=size, y=mmlt, col sep=comma] {ABij_q2_data.csv};
	\addplot table [x=size, y=packed, col sep=comma] {ABij_q2_data.csv};
	\addplot [mark=pentagon*] table [x=size, y=g++, col sep=comma] {ABij_q2_data.csv};
	\addplot [greenclr, mark=diamond*] table [x=size, y=icc, col sep=comma] {ABij_q2_data.csv};
	\end{axis}
	
	\begin{axis}[
	    name=atbtq2,
	    at={($(abtq2.south east)+(-0.5cm,0cm)$)},
	    xshift=1.2cm,
		height=3.5cm,
		width=4.9cm,
		xticklabels={1k, 2k, 4k, 16k, 32k},
		xtick={1,2,3,4,5},
		xlabel={Matrix Order},
		xlabel style={ yshift=1mm, },
		ymax=10,
		ylabel style={ yshift=-6mm, },
		legend style={at={(0.5,-0.23)},
      anchor=north,legend columns=-1,/tikz/every even column/.append style={column sep=0.2cm}},
	]
	\addplot table [x=size, y=mmlt, col sep=comma] {AtBij_q2_data.csv};
	\addplot table [x=size, y=packed, col sep=comma] {AtBij_q2_data.csv};
	\addplot [mark=pentagon*] table [x=size, y=g++, col sep=comma] {AtBij_q2_data.csv};
	\addplot [greenclr, mark=diamond*] table [x=size, y=icc, col sep=comma] {AtBij_q2_data.csv};
	\end{axis}

	\begin{axis}[
	    name=abq3,
	    at={($(abq2.south west)+(0cm,-3cm)$)},
		height=3.5cm,
		width=4.9cm,
		xticklabels={1k, 2k, 4k, 16k, 32k},
		xtick={1,2,3,4,5},
		xlabel={Matrix Order},
		xlabel style={ yshift=1mm, },
		ymax=15,
		ylabel={SPR},
		ylabel style={ yshift=-6mm, },
		legend style={at={(2.41,-0.33)},
      anchor=north,legend columns=-1,/tikz/every even column/.append style={column sep=0.2cm}},
	]
	\addplot table [x=size, y=mmlt, col sep=comma] {ABi_q3_data.csv};
	\addplot table [x=size, y=packed, col sep=comma] {ABi_q3_data.csv};
	\addplot [mark=pentagon*] table [x=size, y=g++, col sep=comma] {ABi_q3_data.csv};
	\addplot [greenclr, mark=diamond*] table [x=size, y=icc, col sep=comma] {ABi_q3_data.csv};
	\end{axis}
	
	\begin{axis}[
	    name=atbq3,
	    at={($(abq3.south east)+(-0.5cm,0cm)$)},
	    xshift=1.2cm,
		height=3.5cm,
		width=4.9cm,
		xticklabels={1k, 2k, 4k, 16k, 32k},
		xtick={1,2,3,4,5},
		xlabel={Matrix Order},
		xlabel style={ yshift=1mm, },
		ymax=15,
		ylabel style={ yshift=-6mm, },
		legend style={at={(0.5,-0.23)},
      anchor=north,legend columns=-1,/tikz/every even column/.append style={column sep=0.2cm}},
	]
	\addplot table [x=size, y=mmlt, col sep=comma] {AtBi_q3_data.csv};
	\addplot table [x=size, y=packed, col sep=comma] {AtBi_q3_data.csv};
	\addplot [mark=pentagon*] table [x=size, y=g++, col sep=comma] {AtBi_q3_data.csv};
	\addplot [greenclr, mark=diamond*] table [x=size, y=icc, col sep=comma] {AtBi_q3_data.csv};
	\end{axis}

	\begin{axis}[
	    name=abtq3,
	    at={($(atbq3.south east)+(-0.5cm,0cm)$)},
	    xshift=1.2cm,
		height=3.5cm,
		width=4.9cm,
		xticklabels={1k, 2k, 4k, 16k, 32k},
		xtick={1,2,3,4,5},
		xlabel={Matrix Order},
		xlabel style={ yshift=1mm, },
		ymax=10,
		ylabel style={ yshift=-6mm, },
		legend style={at={(0.5,-0.23)},
      anchor=north,legend columns=-1,/tikz/every even column/.append style={column sep=0.2cm}},
	]
	\addplot table [x=size, y=mmlt, col sep=comma] {ABij_q3_data.csv};
	\addplot table [x=size, y=packed, col sep=comma] {ABij_q3_data.csv};
	\addplot [mark=pentagon*] table [x=size, y=g++, col sep=comma] {ABij_q3_data.csv};
	\addplot [greenclr, mark=diamond*] table [x=size, y=icc, col sep=comma] {ABij_q3_data.csv};
	\end{axis}
	
	\begin{axis}[
	    name=atbtq3,
	    at={($(abtq3.south east)+(-0.5cm,0cm)$)},
	    xshift=1.2cm,
		height=3.5cm,
		width=4.9cm,
		xticklabels={1k, 2k, 4k, 16k, 32k},
		xtick={1,2,3,4,5},
		xlabel={Matrix Order},
		xlabel style={ yshift=1mm, },
		ymax=10,
		ylabel style={ yshift=-6mm, },
		legend style={at={(0.5,-0.23)},
      anchor=north,legend columns=-1,/tikz/every even column/.append style={column sep=0.2cm}},
	]
	\addplot table [x=size, y=mmlt, col sep=comma] {AtBij_q3_data.csv};
	\addplot table [x=size, y=packed, col sep=comma] {AtBij_q3_data.csv};
	\addplot [mark=pentagon*] table [x=size, y=g++, col sep=comma] {AtBij_q3_data.csv};
	\addplot [greenclr, mark=diamond*] table [x=size, y=icc, col sep=comma] {AtBij_q3_data.csv};
	\end{axis}
	
\end{tikzpicture}
\vspace*{-5mm}
\caption{General Query Performance with different forms of threshold (Higher is better)}
\label{fig:general query threshold form performance linechart}
\end{figure*}

We ran experiments on general, matrix multiplication-like queries to evaluate the performance of \mmlt{} compared to other existing methods used to execute such tasks. We use the 3 classes of queries shown in Figure~\ref{fig:test_queries} for experiments. The first and second queries are explained in Section~\ref{s:introduction}, corresponding to tasks for calculating revenue in a retail data analysis application, and separating strong signals in a machine learning application, respectively. The third query corresponds to a retail data analysis situation, where when {\tt A[i][k]} records how many units of product k were bought by customer i, and {\tt B[k][j]} records the list price of product k offered by seller j, the query calculates how many products would account for over \$100 in sales if customer i was matched with supplier j. In addition, when the matrices are stored as form $AB$ and $A^TB$, \textcolor{black}{for all queries we test all four ways in which the variable in green in Figure~\ref{fig:test_queries}} can be represented, i.e. a {\tt constant, thres[i], thres[j], thres[i][j]}. We however only report results for experiments using {\tt thres[i]} and  {\tt thres[i][j]} as we found that using {\tt constant, thres[i], thres[j]} resulted in similar performance. For all queries we used square matrices of order 1k, 2k, 4k, 16k, and 32k.

Our baselines are a triple nested loop written in C compiled with either g++ or icc with the -O3 flag. We are unable to use OpenBLAS/MKL as a baseline as it does not support execution for general calculations that are not matrix multiplication. We also performed experiments using a triple nested loop written in JavaScript compiled with Graal, but omitted the results as execution time was at least $7 \times$ slower compared to \mmlt{}. The results are shown in Figure ~\ref{fig:general query performance linechart} and Figure ~\ref{fig:general query threshold form performance linechart}.

For queries in Figure ~\ref{fig:general query performance linechart} \mmlt{} without packing is on average \textcolor{black}{$2.64 \times$} faster than icc, and \textcolor{black}{$9.40 \times$} faster compared to g++. We confirmed that g++ and icc generate vectorized code, and icc additionally does automatic loop tiling. \mmlt{} also achieves higher speedup for larger matrices. This is most likely because the memory access pattern of our framework gets comparatively more efficient the larger the matrices are. We observe similar patterns for queries in Figure ~\ref{fig:general query threshold form performance linechart}, where Amulet is on average $1.82\times$ faster than icc and $4.43\times$ faster compared to g++. We can see that \mmlt{} scales well to the size of the matrices it processes, as SPR stays relatively constant for each matrix size.

On small matrices, \mmlt{} with packing performs similarly or slightly worse than \mmlt{} without packing. This is because for small matrices the time spent packing matrices constitutes a bigger portion of the total execution time, and the memory footprint is smaller, thus reducing the potential benefit of packing, and also because for general queries more time is spent executing more instructions, which decreases the relative effect of saving time by reducing TLB misses. For larger queries, \mmlt{} with packing is around 10--30\% faster compared to \mmlt{} without packing.

\begin{table}[]
\centering
\caption{Performance on non-square matrices}
\label{tab:non-square results}
\def\arraystretch{1.1}%
\begin{threeparttable}
\setlength\tabcolsep{2.0pt}
\begin{tabular}{|P{1cm}|c|>{\color{black}}c|>{\color{black}}c|>{\color{black}}c|>{\color{black}}c|}
\hline
\multicolumn{2}{|c|}{\multirow{2}{*}{\backslashbox{Query}{Method}}} & \multicolumn{2}{|c|}{\mmlt{}} & \multirow{2}{*}{OpenBLAS} & \multirow{2}{*}{icc} \\ \cline{3-4}
\multicolumn{2}{|c|}{} & Non-square & Square &  &  \\ \thickhline
\multirow{4}{*}{Matmul} & $AB$ & $30.18 \pm 2.00$ & 30.78 & $34.67 \pm 0.50$ & $9.55 \pm 0.70$ \\ \cline{2-6}
& $A^TB$ & $29.81 \pm 2.47$ & 30.78 & $34.16 \pm 0.90$ & $8.36 \pm 0.21$ \\ \cline{2-6}
& $AB^T$ & $30.24 \pm 0.80$ & 30.86 & $34.06 \pm 0.91$ & $9.72 \pm 0.41$ \\ \cline{2-6}
& $A^TB^T$ & $29.99 \pm 2.29$ & 30.49 & $34.20 \pm 0.89$ & $7.13 \pm 0.54$ \\ \thickhline
\multirow{4}{*}{Query 1} & $AB$ & $7.40 \pm 0.07$ & 7.50 & - & $3.95 \pm 0.04$ \\ \cline{2-6}
& $A^TB$ & $7.39 \pm 0.09$ & 7.49 & - & $4.06 \pm 0.10$ \\ \cline{2-6}
& $AB^T$ & $7.42 \pm 0.06$ & 7.50 & - & $4.54 \pm 0.13$ \\ \cline{2-6}
& $A^TB^T$ & $7.39 \pm 0.09$ & 7.49 & - & $1.06 \pm 0.13$ \\ \thickhline
\multirow{4}{*}{Query 2} & $AB$ & $7.42 \pm 0.08$ & 7.51 & - & $3.75 \pm 0.04$ \\ \cline{2-6}
& $A^TB$ & $7.40 \pm 0.09$ & 7.49 & - & $3.55 \pm 0.05$ \\ \cline{2-6}
& $AB^T$ & $7.43 \pm 0.07$ & 7.51 & - & $4.58 \pm 0.13$ \\ \cline{2-6}
& $A^TB^T$ & $7.41 \pm 0.09$ & 7.50 & - & $1.05 \pm 0.13$ \\ \thickhline
\multirow{4}{*}{Query 3} & $AB$ & $ 12.13 \pm 0.20 $ & 12.38 & - & $5.71 \pm 0.17$ \\ \cline{2-6}
& $A^TB$  & $12.07 \pm 12.15$ & 12.35 & -  & $5.63 \pm 0.07$ \\ \cline{2-6}
& $AB^T$  & $12.15 \pm 0.24$ & 12.38 & - & $6.40 \pm 0.18$ \\ \cline{2-6}
& $A^TB^T$ & $12.05 \pm 0.27$ & 12.35 & - & $2.96 \pm 0.10$ \\ \thickhline
\end{tabular}
\begin{tablenotes}[para,flushright]
     Unit: \textcolor{black}{SPR}
\end{tablenotes}
\end{threeparttable}
\vspace{-0mm}
\end{table}

\subsection{Experiments Using Non-square Matrices}\label{non-square}

We conducted experiments to investigate the effect of using non-square matrices on matrix multiplication-like tasks. Let the dimensions of the two matrices being operated on be expressed as $M \times K$ and $K \times N$. For matrix multiplication and each query shown in Figure~\ref{fig:test_queries}, we measured the \textcolor{black}{SPR of when \mmlt{} with packing, OpenBLAS, and icc} processes non-square matrices where one of $M, N, K$ is $4096c$, another is $4096/c$ and the remaining dimension is $4096$, where $c$ is a constant having a value of 2, 4, or 8. In other words, we vary the ratio of $M, N, K$ while keeping the product of $M, N, K$ constant. There are a total of 18 ratios produced by this setup, and for each query we report the average and standard deviation of the SPR across all ratios, along with the SPR of processing a square matrix of order 4096. Results are shown in Table~\ref{tab:non-square results}. \textcolor{black}{For \mmlt{}, the standard deviation in SPR among different ratios is on average 2.9\% of the mean SPR, which is comparable to OpenBLAS (2.3\%), and lower than icc (4.3\%). Additionally, on average there is a 1.7\% difference between the mean SPR for non-square matrices and the SPR of square matrices.} Thus \mmlt{} can effectively process matrices of varying dimensions without significant variation in performance.

\vspace{0mm}
\subsection{Compiler Overheads}

\begin{table}[]
\centering
\caption{Compile Time Results (Unit: Seconds)}
\label{tab:compile time results}
\def\arraystretch{1.1}%
\begin{tabular}{|c|c|c|c|}
\hline
\backslashbox{Task}{Compiler} & \mmlt{} & g++ & icc\\ \thickhline
Matmul & 0.344 & 0.108 & 0.341 \\ \thickhline
Query 1 & 0.413 & 0.111 & 0.355 \\ \thickhline
Query 2 & 0.418 & 0.113 & 0.360 \\ \thickhline
Query 3 & 0.367 & 0.109 & 0.368 \\ \thickhline
\end{tabular}
\end{table}

To evaluate the overhead of compiling code using \mmlt{}, we measured the end-to-end compilation time using \mmlt{} on code that does standard matrix multiplication, and the three queries shown in Figure~\ref{fig:test_queries}. We use the compilation time of g++ and icc on the same tasks as a baseline. For each task we ran experiments on all four combinations of whether A and B is row-major or column-major, but we only report experiment results for when both A and B is row-major as how the matrices are stored does not make a big difference on compilation time.

Results are shown in Table~\ref{tab:compile time results}. We can see that for \mmlt{}, the overhead of compilation is on the order of hundreds of milliseconds, and is on average $3.5\times$ slower than the compilation time of g++, and is either similar or $1.2\times$ slower than the compilation time of icc.  We can also see that \mmlt{}'s compilation time increases the more complicated the code is. Note that for the same task, the compilation time is constant regardless of the size of the data we operate on. Thus, the fraction of compilation time compared to the total execution time gets smaller as we scale to larger amounts of data.

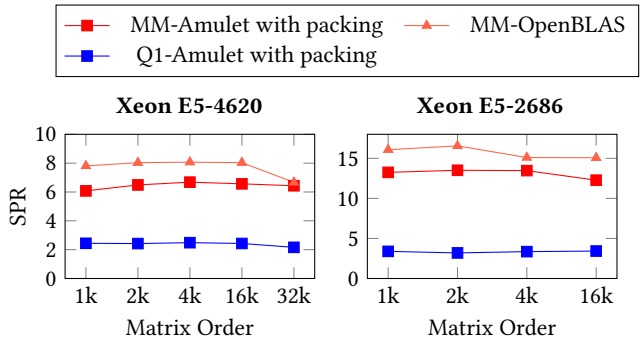
\begin{figure}
\vspace{-0mm}
\hspace{-0.0cm}
\begin{tikzpicture}
    \definecolor{greenclr}{RGB}{34,139,34}
    \definecolor{orangeclr}{RGB}{255,99,71}
    \definecolor{redclr}{RGB}{255,0,0}
    \node[name=ABthresi] at (1.65cm,2.3cm) {\textbf{Xeon E5-4620}};
    \node[name=AtBthresi] at (5.65,2.3cm) {\textbf{Xeon E5-2686}};
	\begin{axis}[
	    name=abq1,
	    at={(0.0cm, 0cm)},
		height=3.5cm,
		width=4.9cm,
		xticklabels={1k, 2k, 4k, 16k, 32k},
		xtick={1,2,3,4,5},
		xlabel={Matrix Order},
		xlabel style={ yshift=1mm, },
		ylabel={SPR},
		ymax=10,
            ymin=0,
		ylabel style={ yshift=-6mm, },
		legend style={legend columns=2, at={(2.3,1.9)},, style={column sep=0.2cm}},
	]
	\addplot [redclr, mark=square*] table [x=size, y=mmlt, col sep=comma] {ingatan_data.csv};
        \addplot [orangeclr, mark=triangle*] table [x=size, y=openblas, col sep=comma] {ingatan_data.csv};
	\addplot [blue, mark=square*] table [x=size, y=q1, col sep=comma] {ingatan_data.csv};
        \legend{MM-\mmlt{} with packing, MM-OpenBLAS,Q1-\mmlt{} with packing}
	\end{axis}

	\begin{axis}[
	    name=abtq1,
	    at={($(abq1.south east)+(-0.5cm,0cm)$)},
	    xshift=1.2cm,
		height=3.5cm,
		width=4.9cm,
		xticklabels={1k, 2k, 4k, 16k, 32k},
		xtick={1,2,3,4,5},
		xlabel={Matrix Order},
		xlabel style={ yshift=1mm, },
		ymax=18,
            ymin=0,
		ylabel style={ yshift=-6mm, },
	]
	\addplot [redclr, mark=square*] table [x=size, y=mmlt, col sep=comma] {aws_data.csv};
        \addplot [orangeclr, mark=triangle*] table [x=size, y=openblas, col sep=comma] {aws_data.csv};
	\addplot [blue, mark=square*] table [x=size, y=q1, col sep=comma] {aws_data.csv};
	\end{axis}
	
\end{tikzpicture}
\vspace*{0mm}
\caption{\textcolor{black}{Query Performance on Two Additional Machines}} 
\label{fig:adaptivity on different machines linechart}
\vspace*{0mm}
\end{figure}

\subsection{\textcolor{black}{Adaptivity to Different Machines}}
\textcolor{black}{
We ran matrix multiplication and query 1 using \mmlt{} with packing on two additional machines. One machine uses a 2.2GHz Xeon E5-4620 processor, while another uses a 2.3GHz Xeon E5-2686 processor. Both machines have 16 256-bit SIMD registers, in contrast to the 32 512-bit SIMD registers supported by the machine used in previous tests. As a baseline, we compare \mmlt{}'s matrix multiplication performance to OpenBLAS. Results are shown in Figure~\ref{fig:adaptivity on different machines linechart}. While g++ and icc results are not shown, we still maintain the relative speedups shown in Section~\ref{s:general code results} on large matrices. Results on matrices with order 32k are not shown for the Xeon E5-2686 processor due to insufficient RAM. \mmlt{} is able to adapt to different machines as it maintains similar relative performance to OpenBLAS (around 20\% slower or less) despite OpenBLAS using different hand-tuned code for each machine.
}

\section{Conclusions}\label{s:conclusions}

We propose \mmlt{}, a compiler that can adaptively optimize matrix multiplication-like tasks written as a program in an imperative programming language. \mmlt{} generates code tailored for each execution environment by using adaptive query processing techniques to test various code parameters at runtime, and optimizes matrix multiplication-like tasks by generating code that performs loop blocking at each level of the memory hierarchy. \mmlt{} achieves speedups on a wide variety of matrix multiplication-like tasks compared to existing compilers that automatically do loop-tiling and code vectorization, and scales well with data size. As a side benefit, on large matrices, the time taken by \mmlt{} to perform matrix multiplication is within 15\% of libraries that are hand-written for each different hardware. We expect that \mmlt{} will improve productivity for common data analysis tasks and facilitate research in the ML community by allowing scientists to write simple code that is also very efficient.

For future work we plan on extending \mmlt{} to support adaptive execution of more types of queries that can be expressed as nested loops, such as the convolutional neural network computations in ~\cite{yang2016systematic} that can be represented as 6-layer nested loops. We also hope to extend \mmlt{} to utilize parallelization, and also execution in a GPU environment.

\begin{acks}
This work was supported by the National Science Foundation under grant IIS-2008295 and by a gift from Oracle.
\end{acks}

\bibliographystyle{ACM-Reference-Format}
\bibliography{references}

\end{document}